\documentclass[12pt,manuscript]{aastex}
\usepackage{xspace}
\RequirePackage{lineno}

\def\del#1{{}}

\def\Fermi{{\em Fermi}\xspace}

\newcommand{\expval}[1]{\left\langle #1 \right\rangle}
\newcommand{\RA}[3]{#1$^{\mathrm{h}}$#2$^{\mathrm{m}}$#3$^{\mathrm{s}}$}
\newcommand{\Dec}[3]{#1$^{\circ}$#2\arcmin#3\arcsec}
\newcommand{\rmn}{\mathrm}
\newcommand{\CR}{\mathrm{CR}}
\newcommand{\dps}{\displaystyle}

\shorttitle{Coma Cluster Observations with VERITAS and \Fermi}
\shortauthors{VERITAS collaboration}

\begin{document}

\title{Constraints on Cosmic Rays, Magnetic Fields, and Dark Matter from Gamma-Ray Observations of the Coma Cluster of Galaxies with VERITAS and \Fermi}
\author{
T.~Arlen\altaffilmark{1},
T.~Aune\altaffilmark{2},
M.~Beilicke\altaffilmark{3},
W.~Benbow\altaffilmark{4},
A.~Bouvier\altaffilmark{2},
J.~H.~Buckley\altaffilmark{3},
V.~Bugaev\altaffilmark{3},
K.~Byrum\altaffilmark{5},
A.~Cannon\altaffilmark{6},
A.~Cesarini\altaffilmark{7},
L.~Ciupik\altaffilmark{8},
E.~Collins-Hughes\altaffilmark{6},
M.~P.~Connolly\altaffilmark{7},
W.~Cui\altaffilmark{9},
R.~Dickherber\altaffilmark{3},
J.~Dumm\altaffilmark{10},
A.~Falcone\altaffilmark{11},
S.~Federici\altaffilmark{12,13},
Q.~Feng\altaffilmark{9},
J.~P.~Finley\altaffilmark{9},
G.~Finnegan\altaffilmark{14},
L.~Fortson\altaffilmark{10},
A.~Furniss\altaffilmark{2},
N.~Galante\altaffilmark{4},
D.~Gall\altaffilmark{15},
S.~Godambe\altaffilmark{14},
S.~Griffin\altaffilmark{16},
J.~Grube\altaffilmark{8},
G.~Gyuk\altaffilmark{8},
J.~Holder\altaffilmark{17},
H.~Huan\altaffilmark{18},
G.~Hughes\altaffilmark{12},
T.~B.~Humensky\altaffilmark{19},
A.~Imran\altaffilmark{20},
P.~Kaaret\altaffilmark{15},
N.~Karlsson\altaffilmark{10},
M.~Kertzman\altaffilmark{21},
Y.~Khassen\altaffilmark{6},
D.~Kieda\altaffilmark{14},
H.~Krawczynski\altaffilmark{3},
F.~Krennrich\altaffilmark{20},
K.~Lee\altaffilmark{3},
A.~S~Madhavan\altaffilmark{20},
G.~Maier\altaffilmark{12},
P.~Majumdar\altaffilmark{1},
S.~McArthur\altaffilmark{3},
A.~McCann\altaffilmark{16},
P.~Moriarty\altaffilmark{22},
R.~Mukherjee\altaffilmark{23},
T.~Nelson\altaffilmark{10},
A.~O'Faol\'{a}in de Bhr\'{o}ithe\altaffilmark{6},
R.~A.~Ong\altaffilmark{1},
M.~Orr\altaffilmark{20},
A.~N.~Otte\altaffilmark{24},
N.~Park\altaffilmark{18},
J.~S.~Perkins\altaffilmark{25,26},
M.~Pohl\altaffilmark{13,12,31}
H.~Prokoph\altaffilmark{12},
J.~Quinn\altaffilmark{6},
K.~Ragan\altaffilmark{16},
L.~C.~Reyes\altaffilmark{27},
P.~T.~Reynolds\altaffilmark{28},
E.~Roache\altaffilmark{4},
J.~Ruppel\altaffilmark{13,12},
D.~B.~Saxon\altaffilmark{17},
M.~Schroedter\altaffilmark{4},
G.~H.~Sembroski\altaffilmark{9},
C.~Skole\altaffilmark{12},
A.~W.~Smith\altaffilmark{14},
I.~Telezhinsky\altaffilmark{13,12},
G.~Te\v{s}i\'{c}\altaffilmark{16},
M.~Theiling\altaffilmark{9},
S.~Thibadeau\altaffilmark{3},
K.~Tsurusaki\altaffilmark{15},
A.~Varlotta\altaffilmark{9},
M.~Vivier\altaffilmark{17},
S.~P.~Wakely\altaffilmark{18},
J.~E.~Ward\altaffilmark{3},
A.~Weinstein\altaffilmark{20},
R.~Welsing\altaffilmark{12},
D.~A.~Williams\altaffilmark{2},
B.~Zitzer\altaffilmark{5}
}

\and
\author{C. Pfrommer\altaffilmark{29,31},
A. Pinzke\altaffilmark{30}}

\altaffiltext{1}{Department of Physics and Astronomy, University of California, Los Angeles, CA 90095, USA}
\altaffiltext{2}{Santa Cruz Institute for Particle Physics and Department of Physics, University of California, Santa Cruz, CA 95064, USA}
\altaffiltext{3}{Department of Physics, Washington University, St. Louis, MO 63130, USA}
\altaffiltext{4}{Fred Lawrence Whipple Observatory, Harvard-Smithsonian Center for Astrophysics, Amado, AZ 85645, USA}
\altaffiltext{5}{Argonne National Laboratory, 9700 S. Cass Avenue, Argonne, IL 60439, USA}
\altaffiltext{6}{School of Physics, University College Dublin, Belfield, Dublin 4, Ireland}
\altaffiltext{7}{School of Physics, National University of Ireland Galway, University Road, Galway, Ireland}
\altaffiltext{8}{Astronomy Department, Adler Planetarium and Astronomy Museum, Chicago, IL 60605, USA}
\altaffiltext{9}{Department of Physics, Purdue University, West Lafayette, IN 47907, USA }
\altaffiltext{10}{School of Physics and Astronomy, University of Minnesota, Minneapolis, MN 55455, USA}
\altaffiltext{11}{Department of Astronomy and Astrophysics, 525 Davey Lab, Pennsylvania State University, University Park, PA 16802, USA}
\altaffiltext{12}{DESY, Platanenallee 6, 15738 Zeuthen, Germany}
\altaffiltext{13}{Institute for Physics and Astronomy, University of Potsdam, 14476 Potsdam,Germany}
\altaffiltext{14}{Department of Physics and Astronomy, University of Utah, Salt Lake City, UT 84112, USA}
\altaffiltext{15}{Department of Physics and Astronomy, University of Iowa, Van Allen Hall, Iowa City, IA 52242, USA}
\altaffiltext{16}{Physics Department, McGill University, Montreal, QC H3A 2T8, Canada}
\altaffiltext{17}{Department of Physics and Astronomy and the Bartol Research Institute, University of Delaware, Newark, DE 19716, USA}
\altaffiltext{18}{Enrico Fermi Institute, University of Chicago, Chicago, IL 60637, USA}
\altaffiltext{19}{Physics Department, Columbia University, New York, NY 10027, USA}
\altaffiltext{20}{Department of Physics and Astronomy, Iowa State University, Ames, IA 50011, USA}
\altaffiltext{21}{Department of Physics and Astronomy, DePauw University, Greencastle, IN 46135-0037, USA}
\altaffiltext{22}{Department of Life and Physical Sciences, Galway-Mayo Institute of Technology, Dublin Road, Galway, Ireland}
\altaffiltext{23}{Department of Physics and Astronomy, Barnard College, Columbia University, NY 10027, USA}
\altaffiltext{24}{School of Physics \& Center for Relativistic Astrophysics, Georgia Institute of Technology, 837 State Street NW, Atlanta, GA 30332-0430}
\altaffiltext{25}{CRESST and Astroparticle Physics Laboratory NASA/GSFC, Greenbelt, MD 20771, USA.}
\altaffiltext{26}{University of Maryland, Baltimore County, 1000 Hilltop Circle, Baltimore, MD 21250, USA.}
\altaffiltext{27}{Physics Department, California Polytechnic State University, San Luis Obispo, CA 94307, USA}
\altaffiltext{28}{Department of Applied Physics and Instrumentation, Cork Institute of Technology, Bishopstown, Cork, Ireland}
\altaffiltext{29}{Heidelberg Institute for Theoretical Studies, Schloss-Wolfsbrunnenweg 35, D-69118 Heidelberg, Germany}
\altaffiltext{30}{Department of Physics, University of California, Santa Barbara, CA 93106, USA}
\altaffiltext{31}{Corresponding authors: M. Pohl, pohlmadq@gmail.com \& C. Pfrommer, christoph.pfrommer@h-its.org}


\begin{abstract}
Observations of radio halos and relics in galaxy clusters indicate efficient electron
acceleration. Protons should likewise be accelerated and, on account of weak energy losses, can
accumulate, suggesting that clusters may also be sources of very high-energy (VHE; $E>100$ GeV)
gamma-ray emission. We report here on VHE gamma-ray observations of the Coma galaxy cluster with
the VERITAS array of imaging Cherenkov telescopes, with complementing \Fermi-LAT observations at
GeV energies. No significant gamma-ray emission from the Coma cluster was detected. Integral flux
upper limits at the 99\% confidence level were measured to be on the order of $(2-5)\times
10^{-8}\ {\rm ph.\,m^{-2}\,s^{-1}}$ (VERITAS, $>220\ {\rm GeV}$) and $\sim 2\times 10^{-6}\ {\rm
  ph.\,m^{-2}\, s^{-1}}$ (\Fermi, $1-3\ {\rm GeV}$), respectively. We use the gamma-ray upper
limits to constrain CRs and magnetic fields in Coma. Using an analytical approach, the
CR-to-thermal pressure ratio is constrained to be $< 16\%$ from VERITAS data and $< 1.7\%$ from
\Fermi data (averaged within the virial radius). {These upper limits are starting to
  constrain the CR physics in self-consistent cosmological cluster simulations and cap the maximum
  CR acceleration efficiency at structure formation shocks to be $<50\%$. Alternatively, this may
  argue for non-negligible CR transport processes such as CR streaming and diffusion into the
  outer cluster regions. } Assuming that the radio-emitting electrons of the Coma halo result from
hadronic CR interactions, the observations imply a lower limit on the central magnetic field in
Coma of $\sim (2 - 5.5)\,\mu{\rm G}$, depending on the radial magnetic-field profile and on the
gamma-ray spectral index. Since these values are below
those inferred by Faraday rotation measurements in Coma (for most of the parameter space),
this {renders} the hadronic model a
very plausible explanation of the Coma radio halo. Finally, since galaxy clusters are dark-matter
(DM) dominated, the VERITAS upper limits have been used to place constraints on the
thermally-averaged product of the total self-annihilation cross section and the relative velocity
of the DM particles, $\expval{\sigma v}$.
\end{abstract}

%
%


\section{Introduction}
Clusters of galaxies are the largest virialized objects in the Universe, with typical sizes of a
few Mpc and masses on the order of $10^{14}$ to $10^{15} M_{\odot}$. According to the currently
favored hierarchical model of cosmic structure formation, larger objects formed through successive
mergers of smaller objects with galaxy clusters sitting on top of this mass hierarchy
\citep[see][for a review]{article:Voit:2005}. Most of the mass ($\sim$80\%) in a cluster is dark
matter (DM), as indicated by galaxy dynamics and gravitational lensing
\citep{article:DiaferioSchindlerDolag:2008}. Baryonic gas making up the intra-cluster medium (ICM)
contributes about 15\% of the total cluster mass and individual galaxies account for the remainder
(about 5\%). The ICM gas mass also comprises a significant fraction of the observable (baryonic)
matter in the Universe.

The ICM is a hot ($T\sim 10^{8}$ K) plasma emitting thermal bremsstrahlung in the soft X-ray regime
\citep[see, e.g.,][]{article:Petrosian:2001}. This plasma has been heated primarily through
collisionless structure-formation shocks that form as a result of the hierarchical merging and
accretion processes. Such shocks and turbulence in the ICM gas in combination with intra-cluster
magnetic fields also provide a means to accelerate particles efficiently \citep[see,
e.g.,][]{article:ColafrancescoBlasi:1998, article:Ryu_etal:2003}. Many clusters feature megaparsec
scale halos of nonthermal radio emission, indicative of a population of relativistic electrons
and magnetic fields permeating the ICM \citep{article:Cassano_etal:2010}. There are two competing
theories to explain radio halos. In the ``hadronic model'', the radio-emitting
electrons and positrons are produced in inelastic collisions of cosmic-ray (CR) ions with the thermal gas of the
ICM \citep{article:Dennison:1980, article:EnsslinPfrommerMiniatiSubramanian:2011}. In the
``re-acceleration model'', a long-lived pool of 100-MeV electrons---previously accelerated by
formation shocks, galactic winds, or jets of active galactic nuclei (AGN)---interacts with plasma
waves that are excited during states of strong ICM turbulence, e.g., after a cluster merger. This
may result in second order Fermi acceleration and may produce energetic electrons ($\sim 10$ GeV)
sufficient to explain the observable radio emission \citep{article:SchlickeiserSieversThiemann:1987,
article:BrunettiLazarian:2010}. Observations of possibly nonthermal emission from clusters
in the extreme ultraviolet \citep[EUV; ][]{article:SarazinLieu:1998} and hard X-rays
\citep{article:RephaeliGruber:2002, article:Fusco-Femiano_etal:2004, article:Eckert_etal:2007} may
provide further indication of relativistic particle populations in clusters, although the
interpretation of these observations as nonthermal diffuse emission has been disputed on the basis
of more sensitive observations \citep[see, e.g.,][]{article:Ajello_etal:2009,
article:Ajello_etal:2010, article:Wik_etal:2009}.

Galaxy clusters have, for many years, been proposed as sources of gamma rays. If shock acceleration
in the ICM is an efficient process, a population of highly relativistic CR protons and heavy ions is
to be expected in the ICM. The main energy-loss mechanism for CR hadrons at high energies is pion
production through the interaction of CRs with nuclei in the ICM. Pions are short lived and decay.
The decay of neutral pions produces gamma rays and the decay of charged pions produces muons, which
then decay to electrons and positrons. Due to the low density of the ICM ($n_{\mathrm{ICM}}\sim
10^{-3}$ cm$^{-3}$), the large size and the volume-filling magnetic fields in the ICM, CR hadrons
will be confined in the cluster on timescales comparable to, or longer than, the Hubble time
\citep[][]{article:Volk_etal:1996, article:Berezinsky_etal:1997} and they can therefore
accumulate. For a given CR distribution function, the hadronically induced gamma-ray flux is
directly proportional to the CR-to-thermal pressure fraction, $X_\CR=\expval{P_{\CR}}/
\expval{P_{\mathrm{th}}}$ \citep[see, e.g.,][]{article:EnsslinPfrommerSpringelJubelgas:2007}, where
the brackets indicates volume averages. A very modest $X_{\CR}$ of a few percent implies an
observable flux of gamma rays \citep[e.g.,][]{article:PfrommerEnsslin:2004b}.

Hydrostatic estimates of cluster masses, which are determined by balancing the thermal pressure
force and the gravitational force, are biased low by the presence of any substantial nonthermal
pressure component, including a CR pressure contribution. Similarly, a substantial CR
pressure can bias the temperature decrement of the cosmic microwave background (CMB) due to the
Sunyaev-Zel'dovich effect in the direction of a galaxy cluster. This could then severely
jeopardize the use of clusters to determine cosmological parameters. Comparing X-ray and optical
potential profiles in the centers of galaxy clusters yields an upper limit of 20-30\% of nonthermal
pressure (that can be composed of CRs, magnetic fields or turbulence) relative to the thermal gas
pressure \citep{article:Churazov_etal:2008, article:Churazov_etal:2010}. An analysis that compares
spatially resolved weak gravitational lensing and hydrostatic X-ray masses for a sample of 18 galaxy
clusters detects a deficit of the hydrostatic mass estimate compared to the lensing mass of $20\%$
at $R_{500}$ -- the radius within which the mean density is 500 times the critical density of the
Universe -- suggesting again a substantial nonthermal pressure contribution on large scales
\citep{article:Mahdavi_etal:2008}. Observing gamma-ray emission is a complementary method of
constraining the pressure contribution of CRs that is most sensitive to the cluster core
region. However, it assumes that the CR component is fully mixed with the ICM and may not allow for
a detection of a two-phase structure of CRs and the thermal ICM. An $X_\CR$ of only a few percent is
required in order to produce a gamma-ray flux observable with the current generation of gamma-ray
telescopes, rendering this technique at least as sensitive as the dynamical and hydrostatic methods
(which are more general in that they are sensitive to any nonthermal pressure component).

Gamma-ray emission can also be produced by Compton up-scattering of ambient photons, for example
CMB photons, on ultra-relativistic electrons. Those electrons can either be secondaries from the
CR interactions mentioned above, or injected into the ICM by powerful cluster members and
further accelerated by diffusive shock acceleration or turbulent reacceleration processes
\citep[][and references therein]{article:SchlickeiserSieversThiemann:1987}.

A third mechanism for gamma-ray production in a galaxy cluster could be self-annihilation of a DM
particle, e.g., a weakly interacting massive particle (WIMP). As already mentioned, about 80\% of
the cluster mass is in the form of dark matter, which makes galaxy clusters interesting targets for
DM searches \citep{article:EvansFerrerSarkar:2004, article:BergstromHooper:2006,
article:PinzkePfrommerBergstrom2009, article:Cuesta_etal:2011} despite their large distances
compared to other common targets for DM searches, such as dwarf spheroidal galaxies
\citep{article:Strigari_etal:2007, article:Acciari_etal:2010, article:Aliu_etal:2009} or the
Galactic Center \citep{article:Kosack_etal:2004, article:Aharonian_etal:2006,
article:Aharonian_etal:2009b, article:Abramowski_etal:2011}. 

While several observations of clusters of galaxies have been made with satellite-borne and
ground-based gamma-ray telescopes, a detection of gamma-ray emission from a cluster has yet to be
made. Observations with EGRET \citep{article:Sreekumar_etal:1996, article:Reimer_etal:2003} and the Large Area Telescope (LAT) on board the
\Fermi Gamma-ray Space Telescope \citep{article:Ackermann_etal:2010} have provided upper limits on
the gamma-ray fluxes (typically $\sim10^{-9}$ ph cm$^{2}$ s$^{-1}$ for \Fermi-LAT observations) for
several galaxy clusters in the MeV to GeV band. Upper limits on the very-high-energy (VHE)
gamma-ray flux from a small sample of clusters, including the Coma cluster, have been provided by
observations with ground-based imaging atmospheric Cherenkov telescopes
\citep[IACTs;][]{article:Perkins_etal:2006, inproc:Perkins_etal:2008, article:Aharonian_etal:2009a,
article:Aleksic_etal:2010,article:Aleksic_etal:2012}.

The Coma cluster of galaxies (ACO 1656) is one of the most thoroughly studied clusters across all
wavelengths \citep{article:Voges_etal:1999}. Located at a distance of about 100 Mpc
\citep[$z=0.023$;][]{article:StrubleRood:1999}, it is one of the closest massive clusters \citep[$M
\sim10^{15}M_{\odot}$;][]{article:Smith:1983, article:Kubo_etal:2008}. It hosts both a giant radio
halo \citep{article:Giovannini_etal:1993,article:Thierbach_etal:2003} and peripheral radio relic,
which appears connected to the radio halo with a ``diffuse'' bridge \citep[see discussion
in][]{article:BrownRudnick:2010}. It has been suggested \citep{article:Ensslin_etal:1998} and
successively demonstrated by cosmological simulations which model the nonthermal emission
processes \citep{article:PfrommerEnsslinSpringel:2008, article:Pfrommer:2008,
article:Battaglia_etal:2009, article:Skillman_etal:2011}, that the relic could well be an infall
shock. Extended soft thermal X-ray (SXR) emission is evident from the ROSAT all-sky survey in the
0.1 to 2.4 keV band \citep{article:BrielHenryBohringer:1992}. Observations with XMM-Newton
\citep{article:Briel_etal:2001} revealed substructure in the X-ray halo supported by substantial
turbulent pressure of at least $\sim 10 \%$ of the total pressure
\citep{article:Schuecker_etal:2004}. The Coma cluster is a natural candidate for gamma-ray
observations.

In this article, results from the VERITAS observations of the Coma cluster of galaxies are
reported, with complementing analysis of available data from the 
Large Area Telescope (LAT) on board the \Fermi Gamma-ray Space Telescope.
The VERITAS and \Fermi-LAT data have been used to place constraints on cosmic-ray particle populations,
magnetic fields, and dark matter in the cluster. Throughout the analyses, a present day Hubble
constant of $H_{0} = 100h$ km s$^{-1}$ Mpc$^{-1}$ with $h=0.7$ has been used.

\section{VERITAS Observations, Analysis, and Results}
The VERITAS gamma-ray detector \citep{article:Weekes_etal:2002} is an array of four 12 m-diameter
imaging atmospheric Cherenkov telescopes \citep{article:Holder_etal:2006} located at an
altitude of $\sim$1250 m a.s.l. at the Fred Lawrence Whipple Observatory in southern Arizona
(31$^{\circ}$~40\arcmin~30\arcsec~N, 110$^{\circ}$~57\arcmin~07\arcsec~W). Each of the telescopes
is equipped with a 499-pixel camera covering a 3.5$^{\circ}$ field of view. The array, completed in
the fall of 2007, is designed to detect gamma-ray emission from astrophysical objects in the
energy range from 100 GeV to more than 30 TeV. Depending on the zenith angle and quality selection
criteria imposed during the data analysis, the effective energy range may be narrower than that. The energy resolution is $\sim 15$\% and the angular
resolution (68\% containment) is $\sim 0.1^{\circ}$ per event at 1 TeV and slightly larger at low energy. At the time of the Coma
cluster observations, the sensitivity of the array allowed for detection of a point source with a
flux of 1\% of the steady Crab Nebula flux above 300 GeV at the confidence level of five standard
deviations ($5\sigma$) in under 45 hours.\footnote{The integral flux sensitivity above 300 GeV was
improved by about 30\% with the relocation of one telescope in the summer of 2009.}

The Coma cluster was observed with VERITAS between March and May in 2008 with all four telescopes
fully operational. The total exposure amounts to 18.6 hours of quality-selected live time, i.e.,
time periods of astronomical darkness with clear sky conditions and no technical problems with the
array. The center of the cluster was tracked in \emph{wobble} mode, where the expected source
location is offset from the center of the field of view by 0.5 degrees, to allow for simultaneous
background estimation \citep{article:Fomin_etal:1994}. All of the observations were made in a small
range with average zenith angle $\sim 21^{\circ}$.

The data analysis was performed following the standard VERITAS procedures described in
\citet{inproc:Cogan_etal:2007} and \citet{inproc:Daniel_etal:2007}. Prior to event reconstruction
and selection, all shower images are calibrated and cleaned. Showers are then reconstructed for
events with at least two telescopes contributing images that pass the following quality selection
criteria: more than four participating pixels in the camera, number of photoelectrons in the image
larger than 75, and the distance from the image centroid to the center of the camera less
than $1.43^{\circ}$. These quality selection criteria impose an energy threshold\footnote{The energy
threshold is defined as the energy corresponding to the maximum of the product function of the
observed spectrum and the collection area. It does not vary significantly for the different source
scenarios and assumed spectral indices reported in this work.} of about 220 GeV. In addition,
events for which only images from the two closest-spaced telescopes\footnote{In the array
configuration prior to summer 2009, two telescopes  had a separation of only 35 m.} survive quality
selection are rejected, as they introduce an irreducible high background rate due to local muons, 
degrading the instrument sensitivity \citep{article:MaierKnapp:2007}.

Gamma-ray-like events are separated from the CR background by imposing selection criteria (cuts) on
the mean-scaled length and width parameters \citep{article:Aharonian_etal:1997,
article:Krawczynski_etal:2006} calculated from a parametrized moment analysis of the shower images
\citep{inproc:Hillas:1985}. These parameters are averages over the four telescopes weighted with the total amplitude of the images, that measure the image moments width and length scaled with values expected for gamma rays. In this analysis, events with a mean-scaled length in the range 
0.05-1.19 and a mean-scaled width in the range 0.05-1.08 are selected as gamma-ray-like events.
These ranges for the gamma-hadron separation cuts were optimized {\em a priori} for a weak point
source (3\% Crab Nebula flux level) and a differential spectral index of 2.4, using data taken on
the Crab Nebula during the same epoch. Because the VHE gamma-ray spectrum for the Coma cluster is
expected to be a power-law function with an index of about 2.3 \citep{article:PinzkePfrommer:2010},
these cuts are suitable for the analysis of the Coma cluster data set. It is noted that slightly
varying the spectral index ($\pm$ 0.2) does not significantly impact the cuts used for quality
selection and gamma-hadron separation in this work. 

The Coma cluster is a very rich cluster of galaxies with many plausible sites for gamma-ray
emission: the core region, the peripheral radio relic, and individual powerful cluster
member galaxies. VERITAS has a large enough field of view to allow investigation of several of
these scenarios. In this work, the focus has been on the core region and three cluster members. The
core region is treated as either a point
source or a mildly extended source, a uniform disk with intrinsic radius $0.2^{\circ}$ or $0.4^{\circ}$,
similar to the extension of the thermal soft X-ray emission from the core. There
is evidence of a recent merger event between the two central galaxies NGC 4889 and NGC 4874
\citep{article:Tribble:1993}. There is also evidence for an excess of nonthermal X-ray emission
from these galaxies as well as from the galaxy NGC 4921 \citep{article:Neumann_etal:2003}.
Therefore, searches for point-like VHE gamma-ray emission have been conducted at the locations of
these galaxies. The regions of interest considered in this work are summarized in Table
\ref{table:roi}.

\begin{figure*}
\begin{center}
\scalebox{0.45}{\plotone{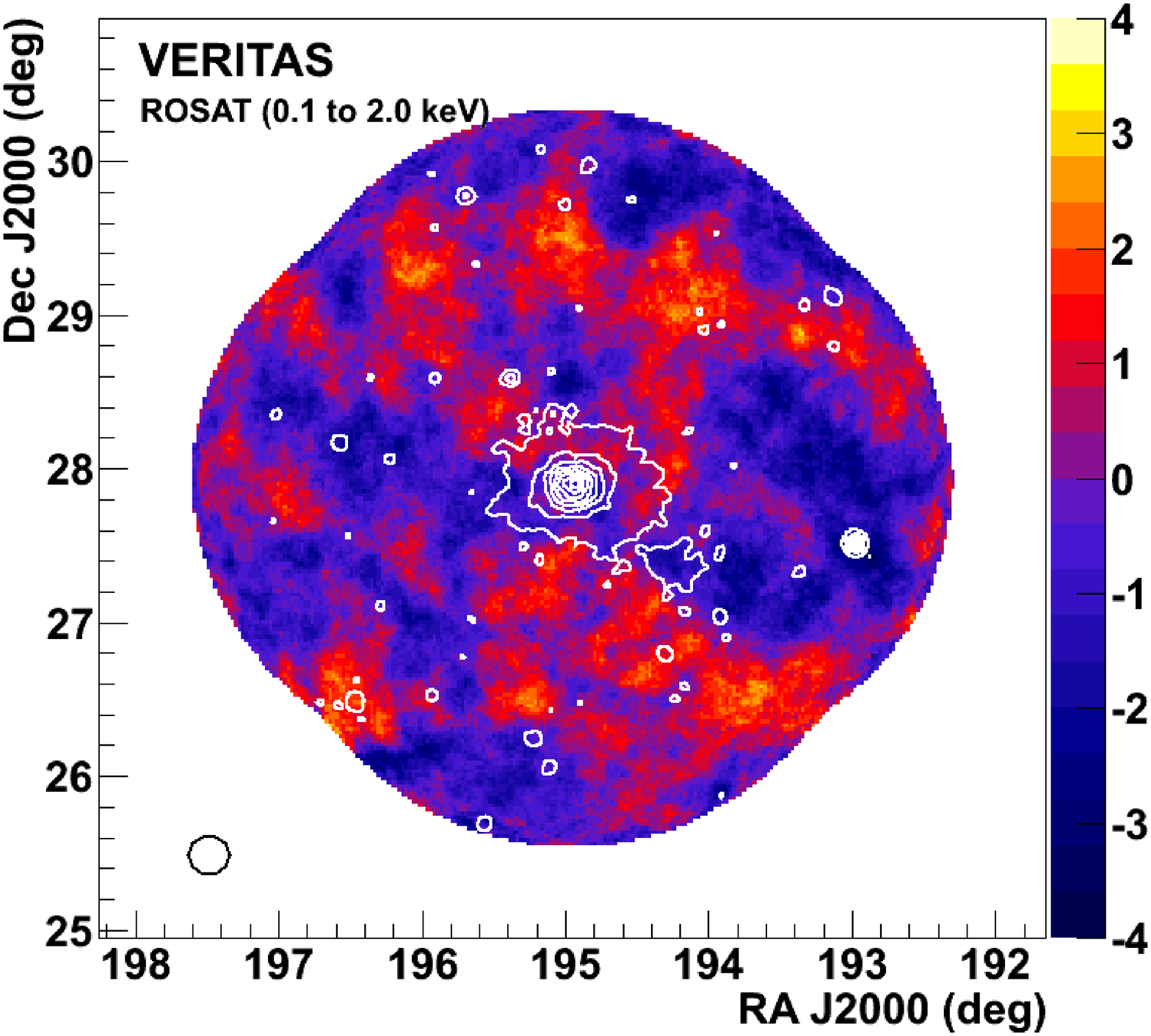}}
\scalebox{0.45}{\plotone{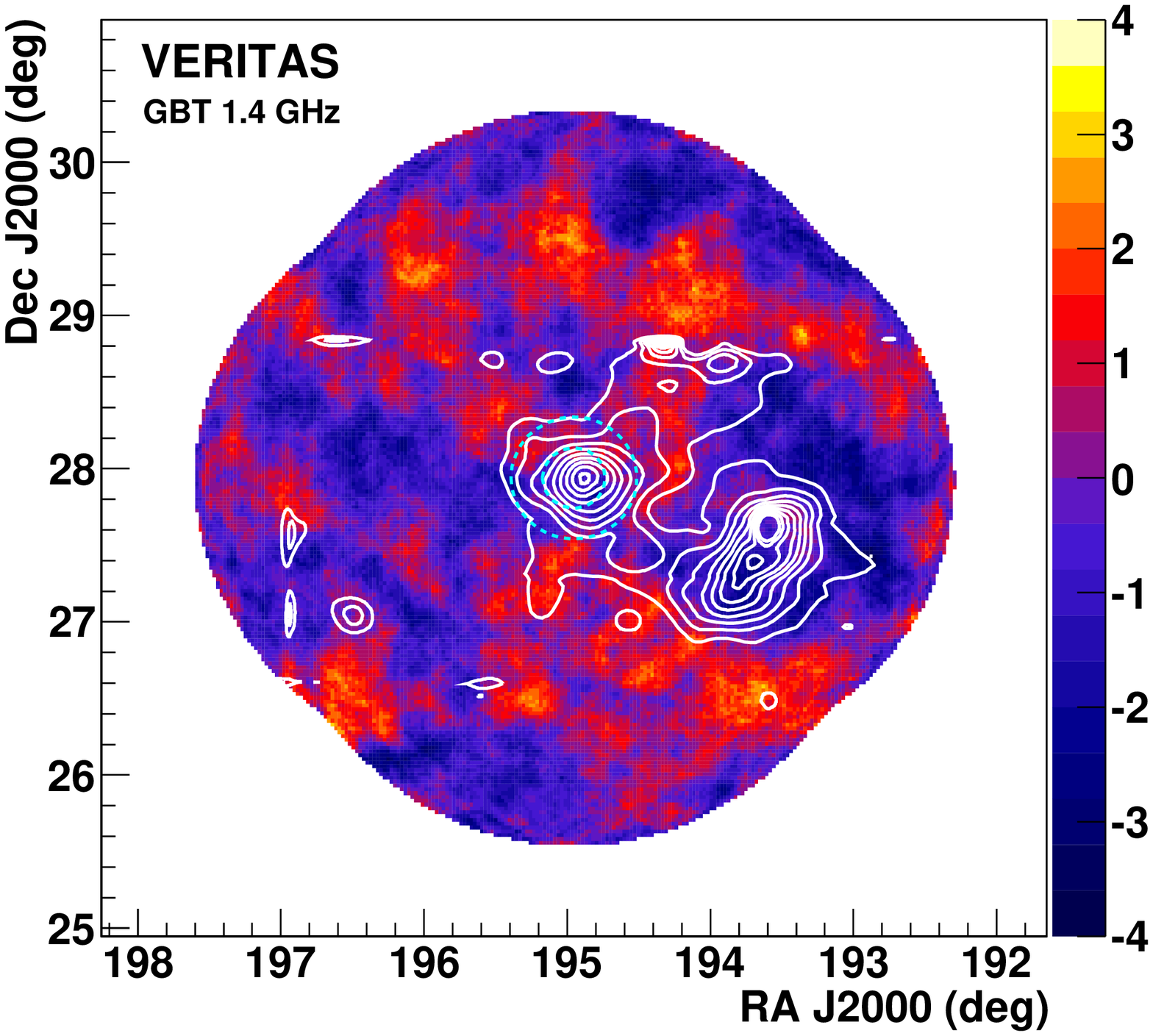}}
\end{center}
\caption{\emph{Left}: Smoothed significance map of the Coma cluster calculated from the excess VHE
gamma-ray events observed with VERITAS over a $4.5^{\circ}\times 4.5^{\circ}$ field of view.
The color scale indicates significance in units of standard deviations. The excess counts were
derived using a ring-background model \citep{article:Aharonian_etal:2001}.
White contours show the X-ray counts per second in the 0.1 to 2 keV energy band
(8 levels from 1 to 16 cts s$^{-1}$ after 3-pixel Gaussian smoothing) from the ROSAT
all-sky survey \citep{article:BrielHenryBohringer:1992}. \emph{Right}: Same as above but with
overlaid contours (20-180 mJ in 20 mJ steps) from GBT radio observations at 1.4 GHz \citep{article:BrownRudnick:2010}, where
strong point sources have been subtracted. Also shown are the $0.2^{\circ}$ and $0.4^{\circ}$ radii
(dashed cyan) considered for the extended-source analyses presented here.}
\label{fig:skymaps}
\end{figure*}

The ring-background model \citep{article:Aharonian_etal:2001} is used to estimate the background
due to CRs misinterpreted as gamma rays (the cuts described above reject more than 99\% of all
CRs). The total number of events in a given region of interest is then compared to the estimated
background from the off-source region scaled by the ratio of the solid angles to produce a final
excess or deficit. The VHE gamma-ray significance is then calculated according to Formula 17 in
\citet{article:LiMa:1983}. Significance skymaps over the VERITAS field of view produced with a
$0.2^{\circ}$ integration radius are shown in Figure \ref{fig:skymaps} with overlaid X-ray and radio
contours from the ROSAT all-sky survey \citep{article:BrielHenryBohringer:1992}  and GBT 1.4 GHz
observations \citep{article:BrownRudnick:2010} respectively.

\begin{figure*}
\begin{center}
\scalebox{1.0}{\plotone{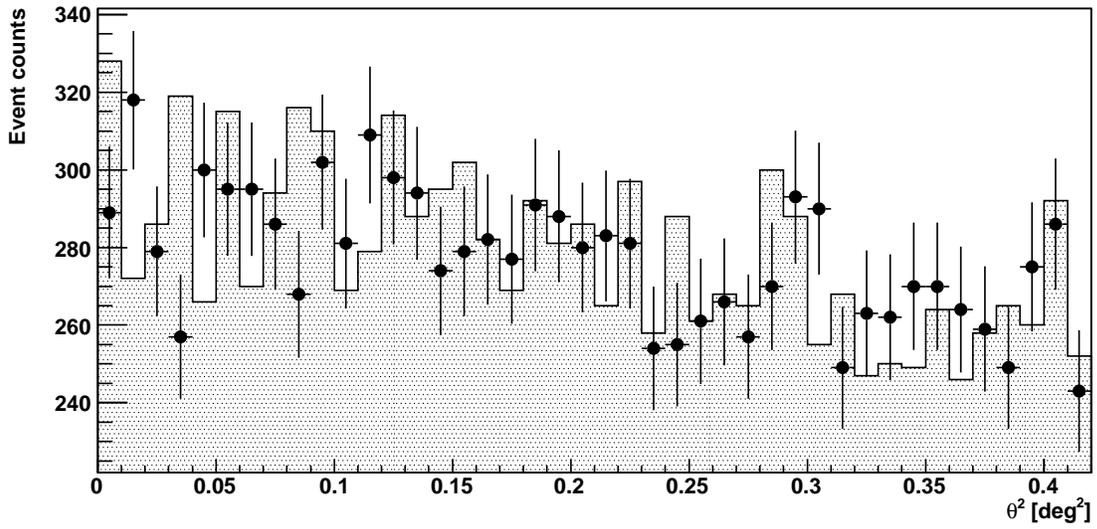}}
\end{center}
\caption{$\theta^{2}$ distribution from VERITAS observations of the Coma cluster of galaxies. The
points with error bars represent the ON-source data sample and the filled area is the background
estimation based on the OFF-source regions. Each bin represents an annulus around the Coma cluster
core position and the annuli are all of equal area. The data were derived from the ring-background
model using a $0.2^{\circ}$ integration radius.}
\label{fig:thetasq}
\end{figure*}

\begin{figure*}
\begin{center}
\scalebox{0.75}{\plotone{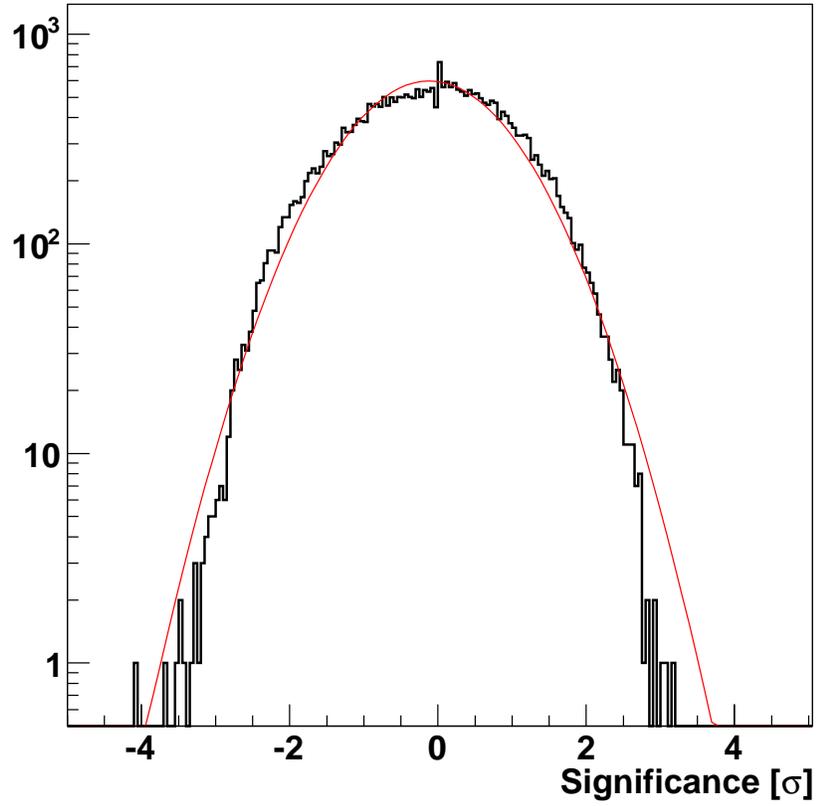}}
\end{center}
\caption{Distribution of significances for Figure \ref{fig:skymaps} and an integration radius of
$0.2^{\circ}$. The curve is a Gaussian fit to the data, with mean $\mu=-0.11\pm 0.0059$ and
standard deviation $\sigma=1.01\pm 0.003$, which is consistent with the absence of gamma-ray
sources in the field of view.}
\label{fig:sigdist}
\end{figure*}

Depending on the assumed extent of the source and the point-spread function, we can define an
ON region, into which a defined fraction of the source photons should fall. No significant excess of VHE gamma rays from the Coma cluster was detected with VERITAS, as
illustrated by the $\theta^{2}$ distribution shown in Figure \ref{fig:thetasq}, in which source events would pile up at small values of $\theta^2$ for a point source and fall into a somewhat wider range of $\theta^2$ values for an extended source. The $\theta^{2}$
distribution is a plot of event density versus the square of the angular separation from a given location. It permits a comparison of the ON-source event distribution with that of other locations, in this case a ring-shaped region, into which only background events should fall, the so-called OFF-source region. The $\theta^{2}$ distribution extends out to 0.42 square degrees to cover both the case of point-like and extended
emission from the core of the Coma cluster. The $\theta^{2}$ distributions for the member galaxies
also considered in this work are very similar to that in Figure \ref{fig:thetasq} and show no
excess of gamma rays. A 99\% confidence level upper limit is calculated for each region of
interest using events from the ON-source and OFF-source regions and the method described by \citet{article:Rolke_etal:2005} assuming a
Gaussian-distributed background. A lower bound of zero is imposed on the gamma-ray flux from the Coma cluster, which prevents artificially low flux upper limits in the case that the best-fit source flux is formally negative. Figure \ref{fig:sigdist} shows the
distribution of significances over the VERITAS skymap, which is well fit by a Gaussian with a mean
close to zero and a standard deviation within a few percent of unity.

Table \ref{table:results} lists the upper limits for the selected regions of interest shown in
Table \ref{table:roi}. These upper-limit calculations depend on the gamma-ray spectrum, which in
this work is assumed to be a power law in energy, $dN/dE\propto E^{-\alpha}$, where the spectral index
$\alpha$ was allowed to have a value of 2.1, 2.3, or 2.5.

\section{\Fermi-LAT Analysis and Results}
LAT on board \Fermi has observed the Coma cluster in all-sky survey mode since its launch in June 2008. \Fermi-LAT is sensitive to gamma rays in the 20 MeV to $\sim300$ GeV energy range and is complementary to the VERITAS observations. \citet{article:Ackermann_etal:2010} reported on the search for gamma-ray emission from thirty-three galaxy clusters in the data from the first 18 months, including the Coma cluster, for which an upper limit of $4.58\times10^{-9}$ ph cm$^{-2}$ s$^{-1}$ in the 0.2 to 100 GeV energy band was reported. This limit is expected to improve as the exposure is increased. In this work an updated analysis is presented as a complement to the VERITAS results which includes data taken between August 5, 2008 and April 17, 2012.

The LAT-data analysis of this work follows the same procedure as described in detail in \citet{2012ApJS..199...31N} and was performed with the Fermi Science Tools version 9.23.1. To only include events with high probability of being photons, the P7SOURCE class and the corresponding P7SOURCE\_V6 instrument-response functions were used throughout this work.

A zenith-angle cut of 100$^\circ$ was applied to eliminate albedo gamma rays from the Earth's limb, excluding time intervals during which any part of the region of interest (ROI) was outside the field of view. In addition, time intervals were removed during which the observatory was transiting the Southern Atlantic Anomaly or the rocking angle exceeded 52$^\circ$.

The ROI is defined to be a square region of the sky measuring $14^{\circ}$ on a side and centered on $\alpha_{J2000}=194.953$ and $\delta_{J2000}=27.9806$, the nominal position of the Coma cluster.

Only photons with reconstructed energy greater than 1 GeV are considered, for which the 68\%-containment radius of the point-spread function (PSF) is narrower than $\sim0.8^{\circ}$. The Fermi-LAT collaboration estimates the systematic uncertainties on the effective area at 10 GeV to be around 10\% \footnote{http://fermi.gsfc.nasa.gov/ssc/data/analysis/LAT\_caveats.html}.

The background emission in the ROI was modeled using fourteen point sources listed in the second LAT source catalog \citep{2012ApJS..199...31N}, the LAT standard Galactic diffuse emission component (\texttt{gal\_2yearp7v6\_v0.fit}), and the corresponding isotropic template (\texttt{iso\_p7v6source.txt}) that accounts for extragalactic emission and residual cosmic-ray contamination. Due to the large tails of the PSF at low energy, further fourteen point sources, lying $\sim4^\circ$ outside the ROI, were included in the source model.

The energy spectra of twenty-four sources are described by a power law. The remaining four sources\footnote{2FGLJ1303.1+2435, 2FGLJ1310.6+3222, 2FGLJ1226.0+2953, and 2FGLJ1224.9+2122}, being bright sources, are modeled with additional degrees of freedom using the log-normal representation, which is typically used for modeling Blazar spectra.

The analysis is performed in three energy bins: 1--3 GeV, 3--10 GeV, and 10--30 GeV.
To find the best fit spectral parameters, a binned maximum-likelihood analysis \citep{1996ApJ...461..396M} is performed for each energy bin on a map with $0.1^{\circ}$ pixel size in gnomonic (TAN) projection, covering the entire ROI. To determine the significance of the sources, and in particular that of the Coma cluster, the analysis tool uses the likelihood-ratio test statistic \citep{1996ApJ...461..396M} defined as,
\begin{equation}
{\rm TS}=-2\left(\ln L_0-\ln L\right),
\end{equation}
where $L_0$ is the maximum likelihood value for the null hypothesis and $L$ is the maximum likelihood with the additional source at a given position on the sky.

In the likelihood analysis the spatial parameters of the sources were kept fixed at the values given in the catalog, whereas the spectral parameters of the point sources in the ROI, along with the normalization of the diffuse components, were allowed to freely vary. We analyzed three cases in which the gamma-ray emission from the Coma cluster was assumed to follow a power-law spectrum with a photon index $\alpha=2.1$, 2.3, and 2.5. The spectral indices of all point sources were permitted to freely vary between $\alpha=0$ and $\alpha=5$. We considered the emission as being caused both by a point-like and a spatially extended source (a uniform disk) with radius $r=0.2^\circ$ or $r=0.4^\circ$, as in the VERITAS analysis.

No significant gamma-ray signal was detected. For one free parameter, the flux from the Coma cluster, the detection significance is computed as the square root of the test statistic (TS follows a $\chi_1^2$ distribution). The highest test statistic was obtained for the high-energy band, where TS~$\sim0.8$ for the point source model, TS~$\sim0.7$ for the disk model with $r=0.2^{\circ}$, and TS~$\sim2$ for the disk model with $r=0.4^{\circ}$.

We therefore used the profile likelihood method \citep{article:Rolke_etal:2005} to derive flux upper limits at the 99\% confidence level in the energy range 1--30~GeV, assuming both an unresolved, point-like or spatially extended emission, as shown in Table~\ref{table:fermi}. 

\section{Gamma Ray Emission from Cosmic Rays}
We decided to adopt a multifaceted approach to constrain the CR-to-thermal pressure distribution in
the Coma cluster using the upper limits derived from the VERITAS and \Fermi-LAT data in this work.
This approach includes (1) a simplified multi-frequency analytical model that assumes a constant
CR-to-thermal energy density and a power-law spectrum in momentum, (2) an analytic model derived
from cosmological hydrodynamical simulations of the formation of galaxy clusters, and (3) a
model that uses the observed intensity profile
of the giant radio halo in Coma to place a lower limit on
the expected gamma-ray flux in the hadronic model -- where the radio-emitting electrons are
secondaries from CR interactions and which is independent of the magnetic field distribution.  This
last approach translates into a minimum CR pressure which, if challenged by tight gamma-ray
limits/detections, permits scrutiny of the hadronic interaction model of the formation of giant
radio halos. Alternatively, realizing a spatial CR distribution that is consistent with the flux
upper limits, and requiring the model to match the observed radio data, enables us to derive a
lower limit on the magnetic field distribution. We stress again that this approach assumes the
validity of the hadronic interaction model. Modeling the CR distribution through different
techniques enables us to bracket our lack of understanding about the underlying plasma physics that
shapes the CR distribution hence to reflect the Bayesian priors that are imposed on the modeling
\citep[see][for a discussion]{article:PinzkePfrommerBergstrom}.

\subsection{Simplified analytical model}
\label{sec:simple}

We start by adopting a simplified analytical model that assumes a power-law CR spectrum and a
constant CR-to-thermal pressure ratio, i.e., we adopt the isobaric model of CRs following
the approach of \citet{article:PfrommerEnsslin:2004b}. To be independent of additional assumptions
and in line with earlier work in the literature, we do not impose a low-momentum cutoff, $q$, on the
CR distribution function, i.e., we adopt $q=0$. Since, {\em a priori}, the CR spectral index is
unconstrained,\footnote{The hadronic interaction physics guarantees that the CR spectral index
coincides with that of the resulting pion-decay gamma-ray emission at energies $E\gg 1$~GeV that are well above the pion bump \citep[see discussion
in][]{article:PfrommerEnsslin:2004b}.} we vary it in the range $2.1<\alpha<2.5$, which is
compatible with the radio spectral index of the giant radio halo of the Coma cluster after
accounting for the spectral steepening at frequencies $\nu\sim5~\rmn{GHz}$ due to the
Sunyaev-Zel'dovich effect \citep{article:Ensslin:2002,
article:PfrommerEnsslin:2004b}.\footnote{Assuming a magnetic field of 1 $\mu$G, the CR protons
responsible for the GHz radio emitting electrons have an energy of $\sim100$ GeV and are $\sim$ 20
times less energetic than those CR protons responsible for 200-GeV gamma-ray emission.} To model
the thermal pressure, we adopt the electron density profile for the Coma cluster that has been
inferred from ROSAT X-ray observations \citep{article:BrielHenryBohringer:1992} and use a constant
temperature of $kT= 8.25$~keV throughout the virial region.

Table \ref{table:constraints_simple} shows the resulting constraints on the CR-to-thermal
pressure ratio, $X_{\CR} = \expval{P_{\CR}}/\expval{P_\rmn{th}}$, averaged within the virial radius,
$R_\rmn{vir}=2.2$~Mpc, that we define as the radius of a sphere enclosing a mean density that is 200
times the critical density of the Universe. Constraints on $X_\CR$ with 
VERITAS flux upper limits (99\% CL) strongly depend on $\alpha$.
This is due to the comparably large energy range from GeV
energies (that dominate the CR pressure, provided $\alpha>2$ and the CR population has a
nonrelativistic low-momentum cutoff, i.e., $q<m_{p}c$, where $m_{p}$ is the proton mass) to energies
at 220 GeV, where our quality selection criteria imposed the energy threshold. These gamma-ray
energies correspond to 1.6 TeV CRs -- an energy ratio of more than 3 orders of magnitude, which
explains the sensitivity to small changes in $\alpha$. The flux measurements within 0.2$^{\circ}$
are the most constraining due to a competition between the integrated signal and the background as
the integration radius increases.  This yields limits on $X_\CR$ between 0.048 and 0.43 (for
$\alpha$ varying between 2.1 and 2.5), with a constraint of $X_\CR<0.1$ for $\alpha=2.3$ (close to
the spectral index predicted by the simulations of \citet{article:PinzkePfrommer:2010} around 220
GeV). Constraints on $X_\CR$ with \Fermi-LAT limits (99\%\ CL) depend only weakly on $\alpha$ because GeV-band gamma rays are produced by CRs with energies near the relativistic transition, that dominantly contribute to the CR pressure.
$X_\CR$-constraints with \Fermi-LAT limits are most constraining for an aperture of
0.4$^{\circ}$; despite the slightly weaker flux upper limits in comparison to the smaller radii of
integration, we expect a considerably larger gamma-ray luminosity due to the increasing volume in
this model. The best limit of $X_\CR< 0.012$ is achieved for $\alpha=2.3$, while the limit for
$\alpha=2.1$ is only slightly worse $(X_\CR<0.017)$.

\subsection{Simulation-based approach}
\label{sec:simulation}
We complement the simplified analytical analysis with a more realistic and predictive approach
derived from cosmological hydrodynamical simulations. We adopt the universal spectral and spatial
gamma-ray model developed by \citet{article:PinzkePfrommer:2010} to estimate the emission from
decaying neutral pions which in clusters dominates over the inverse-Compton (IC) emission above
100 MeV. Given a density profile as, e.g., inferred by
cosmological simulations or X-ray observations, the analytic approach models the CR distribution and
the associated radiative emission processes from radio to the gamma-ray band. This formalism was
derived from high-resolution simulations of clusters of galaxies that included radiative
hydrodynamics, star formation and supernova feedback, and it followed the CR physics by tracing the
most important injection and loss processes self-consistently while accounting for the CR pressure
in the equation of motion
\citep{article:PfrommerSpringelEnsslinJubelgas, article:EnsslinPfrommerSpringelJubelgas:2007,
article:JubelgasSpringelEnsslinPfrommer:2008}. The results are in line with earlier numerical
results on some of the overall characteristics of the CR distribution and the associated radiative
emission processes \citep{article:DolagEnsslin:2000, article:MiniatiRyuKangJones:2001,
article:Miniati:2003, article:Pfrommer_etal:2007, article:PfrommerEnsslinSpringel:2008,
article:Pfrommer:2008}.
 
The overall normalization of the CR and gamma-ray distribution scales nonlinearly with the
acceleration efficiency at structure formation shocks. Following recent observations of supernova
remnants \citep{article:Helder_etal:2009} as well as theoretical studies
\citep{article:KangJones:2005}, we adopt an optimistic but nevertheless realistic value of this
parameter and assume that 50\% of the dissipated energy at strong shocks is injected into CRs, with
this efficiency decreasing rapidly for weaker shocks. Since the vast majority of internal formation
shocks (merger and flow shocks) are weak shocks with Mach numbers $M\lesssim3$
\citep[e.g.,][]{article:Ryu_etal:2003}, they do not contribute significantly to the CR population in
clusters. Instead, strong shocks during the formation epoch of clusters and strong accretion shocks
at the present time (at the boundary of voids and filaments/supercluster regions) dominate the
acceleration of CRs which are adiabatically transported through the cluster. Hence, the
model provides a plausible upper limit for the CR contribution from structure formation shocks in
galaxy clusters which can be scaled with the effective acceleration efficiency. Other possible CR
sources, such as AGN and starburst-driven galactic winds have been neglected for simplicity but
could in principle increase the expected gamma-ray yield.

These cosmological simulations only consider advective transport of CRs by {bulk gas flows that
inject a turbulent cascade, leading to} centrally-enhanced density profiles. However, other means
of CR transport such as diffusion and streaming may flatten the CR radial profiles. {The CRs
stream along magnetic field lines in the opposite direction of the CR number density gradient (at
any energy). In the stratified cluster atmosphere, this implies a net flux of CRs towards larger
radii, equalizing the CR number density with time if not counteracted by advective transport. It
has been suggested that advection velocities only dominate over the CR streaming velocities for
periods with trans- and supersonic cluster turbulence during a cluster merger and drop below the
CR streaming velocities for relaxing clusters.  As a consequence, a bimodality of the CR spatial
distribution is expected to result; with merging (relaxed) clusters showing a centrally
concentrated (flat) CR energy density profile
\citep{article:EnsslinPfrommerMiniatiSubramanian:2011}. This translates into a bimodality of the
expected diffuse radio and gamma-ray emission of clusters, since more centrally concentrated CRs
will find higher target densities for hadronic CR proton interactions. As a result of this,
relaxed clusters could have a reduced gamma-ray luminosity by up to a factor of five
\citep{article:EnsslinPfrommerMiniatiSubramanian:2011}.}  Hence, tight upper limits on the
gamma-ray emission can constrain a combination of acceleration physics and transport properties of
CRs.

We adopt the density profile of thermal electrons as discussed in \S \ref{sec:simple} and model the
temperature profile
of the Coma cluster with a constant central temperature of $kT= 8.25$~keV and a characteristic
decline toward the cluster periphery in accordance with a fit to the universal profile derived from
cosmological cluster simulations \citep{article:PinzkePfrommer:2010, article:Pfrommer_etal:2007}
and the behavior of a nearby sample of deep {\em Chandra} cluster data
\citep{article:Vikhlinin_etal:2005}. This enables us to adopt the spatial and spectral distribution
of CRs according to the model by \citet{article:PinzkePfrommer:2010} that neglects the contribution of supernova remnants, AGN,
and cluster galaxies.

\begin{figure*}
\begin{center}
\scalebox{1.0}{\plotone{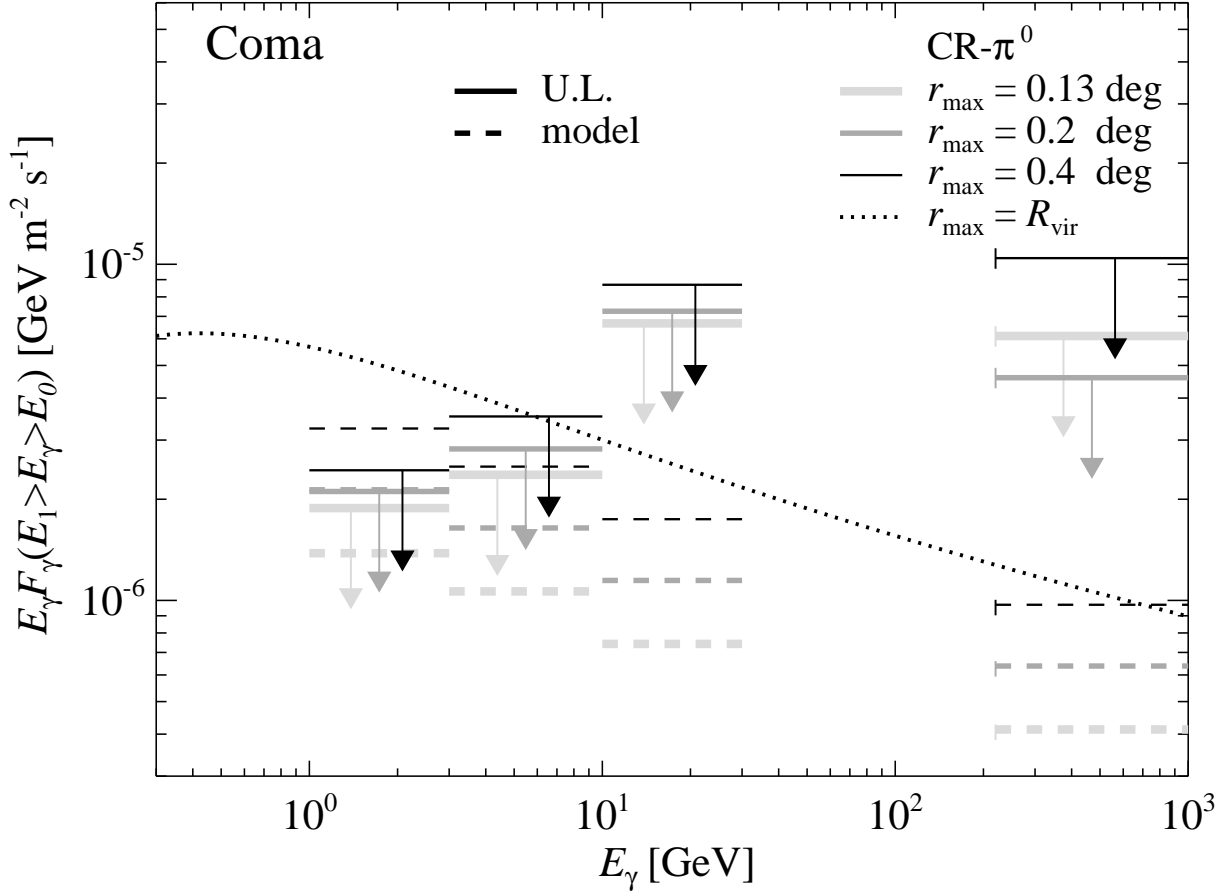}}
\end{center}
\caption{Integral gamma-ray flux upper limits in narrow energy bands $[E_0;E_1]$ 
for \Fermi and VERITAS observations of the Coma
cluster (given in Tables~\ref{table:results} and \ref{table:fermi}) for different integration
radii (arrows with different grey intensities), assuming that $\alpha=2.3$ except 
in the 1--3 GeV
energy interval where $\alpha=2.1$ is adopted. $R_{\rmn{vir}}$ is the virial
radius of the Coma
cluster, corresponding to $1.25^\circ$. These are compared to integrated spectra of the same energy interval and aperture,
assuming the universal gamma-ray spectrum of clusters \citep[lines with different grey
intensities,][]{article:PinzkePfrommer:2010}. To guide the eye, we show the underlying universal
integral energy distribution of pion decay gamma-rays, $E_\gamma F_\gamma(>E_\gamma)$, resulting
from hadronic interactions of CRs and ICM protons (CR-$\pi^0$, dotted). For visualization purposes
all photon fluxes are weighted with the smallest energy in each interval. Note that the \Fermi
limit for the energy interval of 1--3 GeV within the aperture of 0.4$^\circ$ is the most
constraining.}
\label{fig:spectrum}
\end{figure*}

Figure \ref{fig:spectrum} shows the expected integral spectral energy distribution of Coma within
the virial radius (dotted line). This suggests a spectral index of $\alpha=2.1$ in the energy
interval 1-3 GeV and $\alpha=2.3$ for energies probed by VERITAS ($>220$~GeV). Also shown are
integrals of the differential spectrum for finite energy intervals across the angular apertures
tested in this study ({dashed} lines). These model fluxes (summarized in Table
\ref{table:constraints_simple}) are compared to \Fermi and VERITAS flux upper limits for the same
energy intervals. Constraints on $X_\CR$ with the gamma-ray flux limit of \Fermi in the energy
interval 1-3 GeV ($<0.4^\circ$) are most constraining, 
{since that combination of a specific
energy interval and aperture minimizes the ratio of the upper limit to the expected model flux. In
particular, this upper limit is 24\% below} the model predictions that assume an optimistically
large shock-acceleration efficiency and CR transport parameters as laid out above. {Hence this
enables us to constrain a combination of maximum shock acceleration efficiency and CR transport
parameters.}  In our further analysis, we use the most constraining \Fermi-LAT flux limits in the
energy interval 1-3 GeV as well as the gamma-ray flux limits of VERITAS in the energy range above
220 GeV.

\begin{figure*}
\begin{center}
\scalebox{1.0}{\plotone{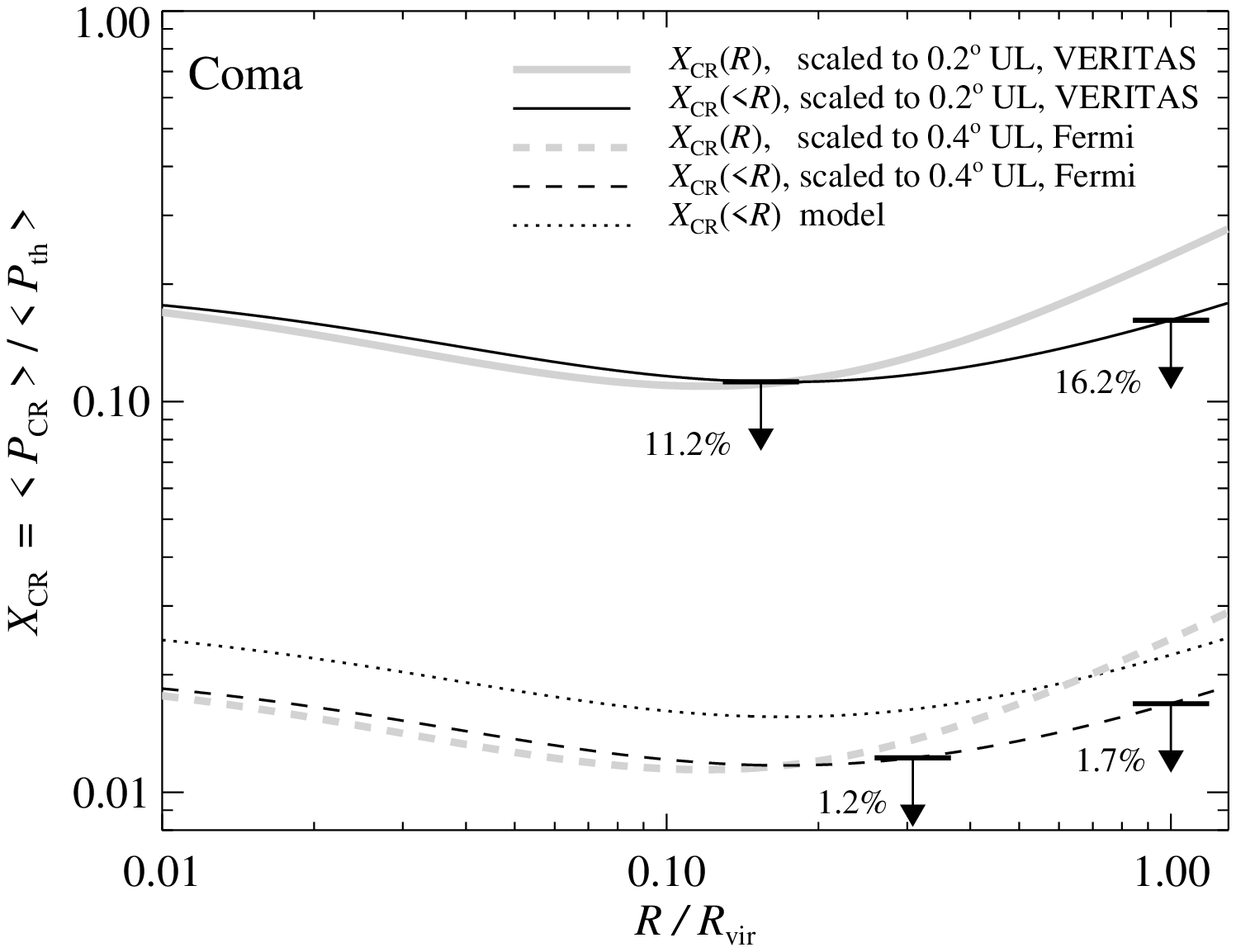}}
\end{center}
\caption{CR-to-thermal pressure ratio, $X_{\mathrm{CR}}=\expval{P_{\CR}}/ \expval{P_{\mathrm{th}}}$, as
a function of radial distance from the center of the Coma cluster. The model predictions, shown by
dashed curves \citep{article:PinzkePfrommer:2010}, have been scaled to match the most
constraining VERITAS upper limits within 0.2$^\circ$ (solid) and \Fermi upper limits within
0.4$^\circ$ (dashed). We compare differential $X_\CR$ profiles (grey) to integrated profiles
$X_{\CR}(<R/R_{\rmn{vir}})=\int_0^R P_\CR\, dV / \int_0^R P_{\rmn{th}}\, dV$ (black) which we use to
compare to the upper limits.}
\label{fig:xcr}
\end{figure*}

Figure \ref{fig:xcr} shows the CR-to-thermal pressure ratio, $X_{\CR} =
\expval{P_{\CR}}/\expval{P_\rmn{th}}$, as a function of radial distance, $R$, from the Coma cluster
center and contained within $R$. All radii are shown in units of the virial radius,
$R_\rmn{vir}=2.2$~Mpc. To compute the CR pressure, we assume a low-momentum cutoff of the CR
distribution at $q = 0.8\,m_{p}c$, where $m_{p}$ is the proton mass. {This is suggested by
cosmological cluster simulations and reflects} the high Coulomb cooling rates at low CR
energies. The CR-to-thermal pressure ratio rises toward the outer regions on account of the higher
efficiency of CR acceleration at the peripheral accretion shocks compared to the weak central flow
shocks. Adiabatic compression of a mixture of CRs and thermal gas disfavors the CR pressure relative
to the thermal pressure on account of the softer equation of state of CRs. The weak increase of
$X_{\CR}$ toward the core is due to the comparably fast thermal cooling of gas.

In the case of VERITAS, for the most constraining regions tested (within an aperture of radius
0.2$^{\circ}$), the predicted CR pressure is a factor of 7.2 below the inferred upper limits of
VERITAS (see Table \ref{table:results} and assuming a spectral index of $\alpha=2.3$ which matches
the simulated one at energies $E_\gamma=200$ GeV). To first order, we can scale the averaged
CR-to-thermal pressure ratio of our model by that factor, keep the spatial behavior and obtain an
integrated limit of the CR-to-thermal pressure ratio of $X_{\CR}<0.112$ within 0.2$^{\circ}$ that
translates to a limit within the cluster virial radius of $X_\CR<0.162$ (solid lines of Figure
\ref{fig:xcr}). This limit is less constraining by 50\% in comparison to the simplified analytical
model, which gives $X_\CR<0.1$. This difference is explained by the concavity of the simulated
spectrum which therefore carries more pressure at GeV energies than a pure power-law spectrum with
$\alpha=2.3$.

{As already aluded to,} the most constraining \Fermi-LAT upper limit in the energy interval 1-3 GeV
($<0.4^\circ$) is a factor of {0.76 smaller} than our model predictions (assuming $\alpha=2.1$
which is very close to the simulated spectral index for the energy range 1-3 GeV). Scaling our
integrated CR-to-thermal pressure profile yields a constraint of $X_{\CR}<0.012$ within
0.4$^{\circ}$ that translates to a limit within the cluster virial radius of $X_\CR<0.017$ (dashed
lines of Figure \ref{fig:xcr}). The $X_\CR$ constraint evaluated within the cluster virial radius is
comparable to the constraint of $X_\CR<0.017$ in our simplified model. Naturally, with the \Fermi-LAT
limits we probe the region around GeV energies that dominate the CR pressure, and we do not expect
any differences to the simplified power-law model in comparison to the universal CR
spectrum with its concave CR spectrum found in the simulations.

\subsection{Minimum gamma-ray flux}
\label{sec:Fmin}

For clusters that host radio halos, we can derive a minimum gamma-ray flux in the hadronic model of
radio halos -- where the radio-emitting electrons are secondaries from CR interactions.  Hadronic
interactions channel about the same power into secondary electrons and $\pi^{0}$-decay gamma rays. A
stationary distribution of CR electrons loses all its energy to synchrotron radiation for strong
magnetic fields ($B \gg B_\rmn{CMB} \simeq (1+z)^2\,3.2 \mu{\rm G}$, 
where $B_\rmn{CMB}$ is the equivalent magnetic
field strength of the CMB so that $B_\rmn{CMB}^2/8\pi$ equals the CMB energy density). Thus the
ratio of gamma-ray to synchrotron flux becomes independent of the spatial distribution of CRs and
thermal gas \citep{article:Voelk:1989, article:Pohl:1994, article:Pfrommer:2008}, in particular with
$\alpha_{\nu}\simeq 1$ as the observed synchrotron spectral index.  Hence we can derive a minimum
gamma-ray flux in the hadronic model
\begin{equation}
\label{eq:Fmin}
F_{\gamma,\rmn{min}} = \frac{\dps A_{\gamma}}{\dps A_{\nu}}\frac{\dps L_{\nu}}{\dps 4\pi D_{\rmn{lum}}^{2}},
\end{equation}
where $L_{\nu}$ is the observed luminosity of the radio mini-halo, $D_{\rmn{lum}}$ denotes the
luminosity distance to the respective cluster, and $A_\gamma$ and $A_\nu$ are dimensional constants
that depend on the hadronic physics of the interaction \citep{article:Pfrommer:2008,
Pfrommer_etal:2008}. Lowering the magnetic field would require an increase in the energy density of
CR electrons to reproduce the observed synchrotron luminosity and thus increase the associated
gamma-ray flux.

To derive a minimum gamma-ray flux that can be compared to the upper limits, we need to determine the
radio flux within the corresponding angular regions. To this end, we fit the
point-source-subtracted, azimuthally-averaged radio-halo profile at 1.38 GHz
\citep{article:Deiss_etal:1997} with a $\beta$-model,
\begin{equation}
\label{beta}
 S_{\nu} (r_{\bot})= S_{0} \left[ 1 + \left( \frac{r_{\bot}}{r_{\rmn{c}}}\right)^{2}\right]^{-3\beta + 1/2},
\end{equation}
where $S_{0} = 1.1 \times 10^{-3}\,\rmn{Jy\,arcmin}^{-2}$, $r_{\rmn{c}} = 450$ kpc, and
$\beta = 0.78$. Within the error bars, this profile is consistent with 326-MHz data taken by
\citet{article:Govoni_etal:2001} when scaled with a radio spectral index of 1.15.

The results for the minimum gamma-ray flux $F_{\gamma,\rmn{min}}(>220~\rmn{GeV})$ and the minimum
CR-to-thermal pressure ratio $X_{\CR,\,\rmn{min}} = X_\CR F_{\gamma,\rmn{min}}/F_{\gamma,\rmn{iso}}$
are shown in Table \ref{table:constraints}, where $F_{\gamma,\rmn{iso}}$ is the gamma-ray flux in
the simplified model introduced in \S\ref{sec:simple}. {Even in the most constraining cases, and
assuming $\alpha\leq 2.3$, these are a factor of $\sim 60$ below the VERITAS upper limits (for
$\alpha=2.1$, $<0.2^{\circ}$) and a factor of $\sim 20$ below the \Fermi-LAT upper limits (for
$\alpha=2.3$, $<0.4^{\circ}$)}. Note that these minimum gamma-ray fluxes are sensitive to the
variation of the CR proton spectral index with energy as a result of, for example,
momentum-dependent diffusion. Assuming a plausible value for the central magnetic field of Coma of
5~$\mu$G \citep{article:Bonafede_etal:2010}, the radio-halo emission at GHz frequencies is dominated
by electrons with energy $E_\rmn{e} \sim 2.5$ GeV (which corresponds to proton energies $E_\rmn{p}
\sim 40$ GeV). Gamma rays with an energy of $200$ GeV are produced by CR protons with an energy of
$E_\rmn{p} \sim 1.6$ TeV -- a factor of 50 higher than those probed by radio halo observations. A
steepening of the CR proton spectral index of 0.2 between 40 GeV and 1.6 TeV would imply a decrease
in the minimum gamma-ray flux by a factor of two.

\subsection{Constraining the Magnetic Field}
\label{sec:B}
In the previous section, we have obtained an absolute lower limit on the gamma-ray emission in
the hadronic model by assuming high magnetic fields, $B\gg B_\rmn{CMB}$. We can turn the
argument around and use our upper limit on the gamma-ray emission (and by extension on the
CR pressure) to infer a lower limit on the magnetic field needed to explain the observed radio
emission. This, again, assumes the validity of the hadronic model of radio halos, in which the
radio-emitting electrons are secondaries from CR interactions. A stronger gamma-ray
constraint will tighten the magnetic-field limit. In case of a conflict with magnetic field
measurements by other methods, e.g., Faraday rotation measure (RM),\footnote{Generally,
Faraday RM analyses of the magnetic field strength by, e.g., background sources observed through
clusters, are degenerate with the magnetic coherence scale and may be biased by the unknown
correlation between magnetic and density fluctuations.} the hadronic model of radio halos would
be challenged. The method we use to constrain the magnetic field inherits a dependence on
the assumed radial scaling which we parametrize as
\begin{equation}
\label{eq:B}
B(r) = B_{0} \,\left(\frac{n_{\rmn{e}}(r)}{n_{\rmn{e}}(0)}\right)^{\alpha_B},
\end{equation}
as suggested by Faraday RM studies and numerical magnetohydrodynamical (MHD) simulations
\citep[][and references therein]{article:Bonafede_etal:2010, article:Bonafede_etal:2011}. Here
$n_{\rmn{e}}$ denotes the Coma electron density profile \citep{article:BrielHenryBohringer:1992}.
In fact, the magnetic field in the Coma cluster is among the best constrained, because its proximity
permits RM observations of seven radio sources located at projected distances of 50 to 1500 kpc from
the cluster center. The best-fit model yields $B_{0} = 4.7^{+0.7}_{-0.8}\,\mu$G and $\alpha_{B} =
0.5^{+0.2}_{-0.1}$ \citep{article:Bonafede_etal:2010}. We aim to constrain the central field
strength, $B_{0}$, and we permit the magnetic decline, $\alpha_{B}$, to vary within a reasonable
range of $\Delta\alpha_{B}=0.2$ as suggested by {those} Faraday RM studies. We proceed as follows:

\begin{enumerate}
\item
Given a model for the magnetic field with $\alpha_B$ and an initial guess for $B_0$, we determine
the profile of the CR-to-thermal pressure ratio, $X_{\CR}(r)$, by matching the hadronically-produced
synchrotron emission to the observed radio-halo emission over the entire extent. To 
this end, we deproject the fit to the surface-brightness profile of Eq. \ref{beta} (using an Abel
integral equation, see Appendix of \citealt{article:PfrommerEnsslin:2004b}) yielding the radio
emissivity,

\begin{equation}
\label{eq:Coma:radio}
j_{\nu} (r) = \frac{S_{0}}{2\pi\, r_{\rmn{c}}}\,
\frac{6\beta - 1}{\left(1 + r^{2}/r_{\rmn{c}}^{2}\right)^{3 \beta}}\,
\mathcal{B}\left(\frac{1}{2}, 3\beta\right)
= j_{\nu,0} \left(1 + r^2/r_{\rmn{c}}^{2}\right)^{-3 \beta},
\end{equation}
where $\mathcal{B}$ denotes the beta function. It is generically true for weak magnetic fields
($B<B_{\rmn{CMB}}$) in the outer parts of the Coma halo that the product $X_{\CR}(r)X_{B}(r)$ (where
$X_B$ denotes the magnetic-to-thermal energy density ratio) has to increase by a factor of about 100
toward the radio-halo periphery to account for the observed extent. If we were to adopt a steeper
magnetic decline such as $\alpha_{B}=0.5$ which produces a flat $X_{B}(r)$, the CR-to-thermal
pressure ratio would have to rise accordingly by a factor of 100.

\item
Given this realization for $X_{\CR}$, we compute the pion-decay gamma-ray surface-brightness
profile, integrate the flux within a radius of $(0.13, 0.2, 0.4)$ degree, and scale the CR profile
in order to match the corresponding VERITAS/\Fermi flux upper limits. This scaling factor,
$X_{\CR,0}$, depends on the CR spectral index, $\alpha$, (assuming a power-law CR population for
simplicity), the radial decline of the magnetic field, $\alpha_{B}$, and our initial guess for
$B_{0}$.

\item
We then solve for $B_{0}$ while matching the observed synchrotron profile and fixing the profile
of $X_{\CR}(r)$ as determined through the previous two steps. Note
that for $B_{0} \gg B_{\rmn{CMB}}$ and a radio spectral index of $\alpha_{\nu}=1$, the solution
would be degenerate since the luminosity of the radio halo scales as
\begin{equation}
\label{eq:Lnu}
L_{\nu} \propto \int dV Q(E)\,\frac{B^{1+\alpha_\nu}}{B^2 + B_{\rmn{CMB}}^2} \to \int dV Q(E),
\end{equation}
where $Q(E)$ denotes the electron source function.

\item Inverse-Compton cooling of CR electrons on CMB photons introduces a characteristic scale of
  $B_{\rmn{CMB}}\simeq 3.2\,\mu$G which imprints as a nonlinearity on the synchrotron emissivity as
  a function of magnetic field strength (see Eq. (\ref{eq:Lnu})). Hence we have to iterate through
  the previous steps until our solution for the minimum magnetic field $B_{0}$ converges.
\end{enumerate}

Table \ref{table:Bmin} shows the resulting lower limit of the central magnetic field ranging from
$B_{0} = 0.5$ to $1.4\,\mu$G in case the of VERITAS and from $B_{0} = 1.4$ to $5.5\,\mu$G in the
case of \Fermi-LAT.\footnote{Note that a central magnetic field of $3\,\mu$G corresponds in the Coma
cluster to a magnetic-to-thermal energy density ratio of $X_B=0.005$.} Since these lower limits on
$B_{0}$ are below the values favored by Faraday RM for {most of the} parameter space spanned by
$\alpha_{B}$ and $\alpha$ {(and never exceed the values for the phenomenological Faraday
RM-inferred $B$-model)}, the hadronic model is a viable explanation of the Coma radio halo. {In fact, the \Fermi-LAT upper limits start to rule out the parameter combination of
$\alpha_{B}\gtrsim 0.7$ and $\alpha \gtrsim 2.5$ for the hadronic model of the Coma radio halo.}
Future gamma-ray observations of the Coma cluster may put more stringent constraints on the
parameters of the hadronic model.

A few remarks are in order. (1) For the VERITAS limits, the hardest
CR spectral indices correspond to the tightest limits on $B_{0}$, because the CR flux is constrained
around 1 TeV and a comparably small fraction of CRs at 100 GeV would be available to produce
radio-emitting electrons. A high magnetic field would be required to match the observed
synchrotron emission. The opposite is true for the \Fermi upper limits at 1 GeV, which
probe CRs around a pivot point of 8 GeV: a soft CR spectral index implies a comparably small
fraction of CRs at 100 GeV and hence a strong magnetic field is needed to match the observed
synchrotron flux. (2) For a steeper
magnetic decline (larger $\alpha_{B}$), the CR number density needs to be larger to match the
observed radio-emission profiles, which would yield a higher gamma-ray flux so that the upper limits
are more constraining. This implies tighter lower limits for $B_{0}$. (3) Interestingly, in all
cases, the 0.4$^{\circ}$-aperture limits are the most constraining. For a given magnetic
realization, a substantially increasing CR-to-thermal pressure profile is needed to match the
observed radio profiles, and therefore that CR realization produces a larger flux within 
0.4$^{\circ}$ in comparison with the simplified CR model ($X_{\CR} = \rmn{const.}$), for which the
0.2$^{\circ}$-aperture limits are more constraining in the case of VERITAS. Physically, the large
CR pressure in the cluster periphery may arise from CR streaming into the large available phase
space in the outer regions.

As a final word, in Table~\ref{table:Bmin} we show the corresponding values for the
CR-to-thermal pressure ratio (at the largest emission radius at 1~Mpc) such that the model
reproduces the observed radio surface-brightness profile.\footnote{Note that in this section, we
determine the radial behavior of $X_\CR$ by adopting a specific model for the magnetic field and
requiring the modeled synchrotron surface-brightness profile to match the observed data of the
Coma radio halo. This is in contrast to the simplified analytical CR model where $X_\CR$ is
constant (\S~\ref{sec:simple}) and to the simulation-based model where $X_\CR(r)$ is derived from
cosmological cluster simulations (\S~\ref{sec:simulation}).} They should be interpreted as upper
limits since they are derived from flux upper limits. For the \Fermi-LAT upper limits, they range from
0.08 to 0.27; hence the $X_{\CR}$ profiles always obey the energy condition, i.e., $P_{\CR} <
P_{\mathrm{th}}$, over the entire range of the radio-halo emission ($< 1$ Mpc).\footnote{See Figure
3 in \citet{article:PfrommerEnsslin:2004a} for the entire parameter range assuming minimum energy
conditions, and \citet{article:PfrommerEnsslin:2004b}, Figure 7 for a parametrization as adopted
in this study. We caution, however, that the minimum-energy condition is violated at the outer
radio-halo boundary for the range of minimum magnetic-field values inferred by this study.} The
corresponding values for $X_\CR$ in the cluster center are smaller than 0.01 for the entire
parameter space probed in this study. We conclude that the hadronic model is not challenged by
current Faraday RM data and is a perfectly viable possibility in explaining the Coma radio-halo
emission.

\section{Emission from Dark Matter Annihilations}
As already mentioned in the introduction, most of the mass in a galaxy cluster is in the form of
DM. While the nature of DM remains unknown, a compelling theoretical candidate is a
WIMP. The self-annihilation of WIMPs can produce either monoenergetic gamma-ray lines or a
continuum of secondary gamma rays that deviates significantly from the power-law spectra observed
from most conventional astrophysical sources, with a sharp cut-off at the WIMP mass. These spectral
features together with the expected difference in the intensity distribution compared to
conventional astrophysical sources allow a clear, indirect detection of DM.

The expected gamma-ray flux due to self-annihilation of WIMPs in a
dark-matter halo is given by 
\begin{equation}
\frac{d\Phi_{\gamma}(\Delta\Omega,E)}{dE}=
\frac{\expval{\sigma v}}{8\pi m_{\chi}^{2}}\,\frac{dN_{\gamma}}{dE}\, J(\Delta\Omega),
\label{eqn:WIMPflux}
\end{equation}
where $\expval{\sigma v}$ is the thermally-averaged product of the total self-annihilation cross
section and the relative WIMP velocity, $m_{\chi}$ is the WIMP mass, $\frac{dN_{\gamma}}{dE}$ is the
differential gamma-ray yield per annihilation\footnote{In this work, we have calculated the
differential gamma-ray yield per annihilation using the Pythia Monte Carlo simulator.},
$\Delta\Omega$ is the observed solid angle, and $J$ is the so-called
astrophysical factor -- a 
factor which determines the DM annihilation rate and depends on the DM distribution.

Given the upper limit on the observed gamma-ray rate, defined as the ratio of the event number detected within the observing time $T_{\mathrm{obs}}$,
$R_{\gamma}(99\%\ \mathrm{CL}) = N_{\gamma}(99\%\ \mathrm{CL}) / T_{\mathrm{obs}}$, we can place
constraints on the WIMP parameter space $(m_{\chi}, \expval{\sigma v})$. Integrating Eq.
(\ref{eqn:WIMPflux}) over energy we find
\begin{equation}
\expval{\sigma v}(99\%\ \mathrm{CL}) <
R_{\gamma}(99\%\ \mathrm{CL})\, \frac{8\pi m_{\chi}^{2}}{J(\Delta\Omega)}\,
\left[\int^{m_{\chi}}_{0} dE\ A_{\mathrm{eff}}\,\frac{dN_\gamma (E)}{dE}\right]^{-1},
\end{equation}
where $A_{\mathrm{eff}}$ is the effective area of the gamma-ray detector.
Because the self-annihilation of a WIMP is a two-body process, the astrophysical factor
$J(\Delta\Omega)$ is the line-of-sight integral of the DM density squared
\begin{equation}
J(\Delta\Omega)=\int_{\Delta\Omega}d\Omega\int d\lambda\ \rho_{\chi}^{2}(\lambda,\Omega),
\end{equation}
where $\lambda$ represents the line of sight. In this work, we have modeled the Coma DM
distribution with a Navarro, Frenk and White (NFW) profile \citep{article:NavarroFrenkWhite:1997},
\begin{equation}
\rho_{\chi}(r)=\rho_{s}\left(\frac{r}{r_{s}}\right)^{-1}\left(1+\frac{r}{r_{s}}\right)^{-2},
\end{equation}
where $r_{s}$ is the scale radius and $\rho_{s}$ is the scale density. Using weak-lensing
measurements of the virial mass in the Coma cluster and the DM
halo mass-concentration relation derived from $N$-body simulation of structure formation
\citep{article:Bullock_etal:2001}, \citet{article:Gavazzi_etal:2009} find, and list in their Table~1 (note that they define $R_{\rm vir}=R_{\rm 100})$, $M_{\rm vir}=M_{\rm 200} =9.7(+6.1/-3.5)\cdot 10^{14}\,
h^{-1}\ {\rm M_\odot}$ and $C_{\rm vir}=C_{\rm 200}=3.5(+1.1/-0.9)$, which we translate into 
the density-profile parameters $r_{s}=0.654$ Mpc and
$\rho_{s}=4.4\times 10^{14}$ M$_{\odot}$/Mpc$^{3}$. Note that the uncertainties
are not necessarily distributed as a Gaussian, and also arise from the choice of dark-matter profile.
According to the latest high-resolution DM-only
simulations of nine rich galaxy clusters, the inner regions of the smooth density profiles are quite
well approximated by the NFW formula \citep{article:Gao_etal:2012}. However, gravitational
interactions of DM with baryons may modify these predictions. This could give rise to either an
increasing inner density slope due to adiabatic contraction of the DM component in response to
cooling baryons in the central regions or a decreasing density slope due to violent baryonic
feedback processes pushing gas out of the center by, e.g., energy injection through AGNs. However,
on scales $r\gtrsim 0.45$~Mpc or more than 20\% of $R_{\rm vir}$ (which are of relevance for the present work), different assumptions
about the inner slope of the smooth DM density profile translate to 
uncertainties in the
resulting astrophysical factor. Table \ref{table:astrofactor} lists the
astrophysical factors calculated for the different VERITAS apertures considered in this work. Table
\ref{table:astrofactor} also lists the astrophysical factor calculated for the background region,
which is used to estimate the gamma-ray contamination from DM annihilation in the background region. As long as the DM contribution to the event number in the background region is negligible, the upper limits derived here directly scale with the
astrophysical factor, ${\rm UL} (<\sigma v>)\propto J^{-1}$. The analysis uses a ring region to estimate the background in a ON region. We have to compute the expected level of gamma-ray emission from DM annihilation in the ring region in order to check that it is negligible with respect to the level of gamma-ray emission from DM annihilation in the ON region. This is equivalent to compute the astrophysical factor of the ON and OFF source region since this quantity is related to the rate of DM annihilations.

\begin{figure*}
\begin{center}
\scalebox{1.0}{\plotone{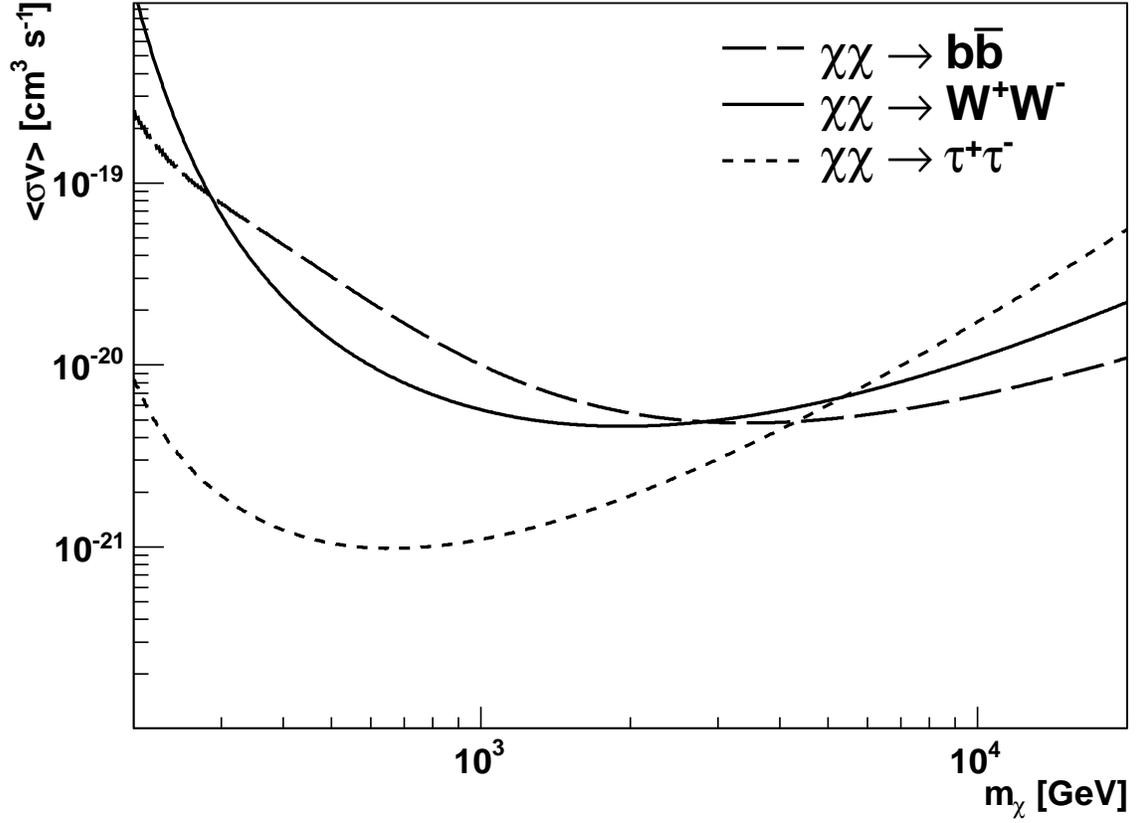}}
\end{center}
\caption{Limits on the DM annihilation cross section $\expval{\sigma v}$ from VERITAS observations
of the Coma cluster as a function of the DM particle mass $m_{\chi}$ derived from the VERITAS
gamma-ray flux upper limits ($0.2^{\circ}$ aperture) presented in this work.}
\label{fig:dm}
\end{figure*}

The resulting exclusion curves in the $(\expval{\sigma v}, m_{\chi})$ parameter space are shown in
Figure \ref{fig:dm} for three different DM self-annihilation channels, W$^{+}$W$^{-}$,
b$\bar{\mathrm{b}}$, and $\tau^{+}\tau^{-}$. Depending on the
DM annihilation channel, the limits are on the order of
$10^{-20}$ to $10^{-21}$ cm$^{3}$ s$^{-1}$. The
minimum for each exclusion curve and corresponding DM particle mass is listed in Table
\ref{table:DMlimits}. We stress that these limits are derived with conservative estimates of the
astrophysical factor $J$. They do not include any boost to the annihilation rate possibly due to DM
substructures populating the Coma halo, which could scale down the present limits by a factor
$O(1000)$ in the most optimistic cases
\citep{article:PinzkePfrommerBergstrom,article:Gao_etal:2012}.

We also note that when the size of the integration region is increased, the limits on
$\expval{\sigma v}$ result from a competition between the gain in the astrophysical factor
$\expval{J}$ and the integrated background. For integration regions larger than 0.2$^{\circ}$ in
radius, the astrophysical factors no longer compensate for the increased number of background
events, and the signal-to-noise ratio deteriorates.

\section{Discussion and Conclusions}
We have reported on the observations of the Coma cluster of galaxies in VHE gamma rays with VERITAS
and complementary observations with the \emph{Fermi}-LAT. VERITAS observed the Coma cluster of
galaxies for a total of 18.6 hours of high-quality live time between March and May in 2008. No
significant excess of gamma rays was detected above an energy threshold of $\sim 220$ GeV. The
\emph{Fermi}-LAT has observed the Coma cluster in all-sky survey mode since its launch in June
2008. We have used all data available from launch to April 2012 for an updated analysis
compared to
published results \citep{article:Ackermann_etal:2010}. Again, no significant excess of gamma rays
was detected. We have used the VERITAS and \emph{Fermi}-LAT data to calculate flux upper limits at
the 99\% confidence level for the cluster core (considered as both a point-like source and a
spatially-extended emission region) and for three member galaxies. The flux upper limits obtained
were then used to constrain properties of the cluster.

We have employed various approaches to constrain the CR population and magnetic field distribution
that are complementary in their assumptions and hence well suited to assessment of the underlying
Bayesian priors in the models. (1) We used a simplified ``isobaric CR model'' that is characterized
by a constant CR-to-thermal pressure fraction and has a power-law momentum spectrum. While this
model is not physically justified {\em a priori}, it is simple and widely used in the literature and
captures some aspects of more elaborate models such as (2) the simulation-based analytical approach
of \citet{article:PinzkePfrommer:2010}. The latter is a ``first-principle approach'' that predicts
the CR distribution spectrally and spatially for a given set of assumptions.
It is powerful since it only requires the density profile as input due to the
approximate universality of the CR distribution (when neglecting CR diffusion and streaming). Note,
however, that inclusion of these CR transport processes may be necessary to explain the radio-halo
bimodality. (3) Finally, we used a pragmatic approach which models the CR and magnetic distributions
in order to reproduce the observed emission profile of the Coma radio halo. While this approach is
also not physically justified, it is powerful because it shows what the ``correct'' model has to
achieve and can point in the direction of the relevant physics.

Within this pragmatic approach, we employ two different methods. Firstly, adopting a high magnetic
field everywhere in the cluster ($B\gg B_\rmn{CMB}$) yields the minimum gamma-ray flux in the
hadronic model of radio halos which we find to be a factor of 20 (60) below the most constraining
flux upper limits of \Fermi-LAT (VERITAS). Secondly, by matching the radio-emission profile (i.e.,
fixing the radial CR profile for a given magnetic field model) and by requiring the pion-decay
gamma-ray flux to match the flux upper limits (i.e., fixing the normalization of the CR
distribution), we obtain lower limits on the magnetic field distribution under consideration. Our
limits for the central magnetic field range from $B_{0} = 0.5$ to $1.4\,\mu$G (for VERITAS flux
limits) and from $B_{0} = 1.4$ to $5.5\,\mu$G (for \Fermi-LAT flux limits). Since all {\em (but one) of}
these lower limits on $B_0$ are below the values favored by Faraday RM, $B_{0} =
4.7^{+0.7}_{-0.8}\,\mu$G \citep{article:Bonafede_etal:2010}, the hadronic model is a very attractive
explanation of the Coma radio halo.  {The \Fermi-LAT upper limits start to rule out the parameter
combination of $\alpha_{B}\gtrsim 0.7$ and $\alpha \gtrsim 2.5$ for the hadronic model of the Coma
radio halo.}

Applying our simplified ``isobaric CR model'' to the most constraining VERITAS limits, we can
constrain the CR-to-thermal pressure ratio, $X_\CR$, to be below 0.048--0.43 (for a CR or
gamma-ray spectral index, $\alpha$, varying between 2.1 and 2.5). We obtain a constraint of
$X_\CR<0.1$ for $\alpha=2.3$, the spectral index predicted by simulations at energies around 220
GeV. This limit is
more constraining by a factor of 1.6 than that of the simulation-based model which gives
$X_\CR<0.16$. This difference is due to the concave form of the simulated spectrum which provides
more pressure at GeV energies in comparison to a pure power-law spectrum of $\alpha=2.3$.

The \Fermi-LAT flux limits constrain $X_\CR$ to be below 0.012--0.017 (for $\alpha$ varying between 2.3
and 2.1), only weakly depending on the assumed CR spectral index. Assuming $\alpha=2.1$, which is
very close to the simulated spectral index for the energy range of 1--3 GeV, we obtain a constraint
which is identical to that from our simulation-based model within the virial radius of
$X_\CR<0.017$. That constraint improves to $X_{\CR}<0.012$ for an aperture of 0.4$^\circ$
corresponding to a physical scale of $R \simeq R_{200}/3 \simeq 660$~kpc.  {These upper limits
are now starting to constrain the CR physics in self-consistent cosmological cluster simulations
and cap the maximum CR acceleration efficiency at structure formation shocks to be
$<50\%$. Alternatively, this may argue for non-negligible CR transport processes such as CR
streaming and diffusion into the outer cluster regions \citep{article:Aleksic_etal:2012}.} These
are encouraging results in that we constrain the CR pressure (of a phase that is fully mixed with
the ICM) to be at most a small fraction ($<0.017$) of the overall pressure. As a result, hydrostatic
cluster masses and the total Comptonization parameter due to the Sunyaev-Zel'dovich effect suffer at
most a very small bias due to CRs.

We have also used the flux upper limits obtained with VERITAS to constrain the thermally-averaged
product of the total self-annihilation cross section and the relative velocity of DM
particles. Modeling the Coma cluster DM halo with a NFW profile we derived limits on
$\expval{\sigma v}$ to be on the order of $10^{-20}$ to $10^{-21}$ cm$^{-3}$ s$^{-1}$ depending on
the chosen aperture. These limits are based on conservative estimates of the astrophysical factor,
where a possible boost to the annihilation rate due to DM substructures in the cluster halo has
been neglected. Including such a boost could scale down the present limits by a factor $O(1000)$ in
the most optimistic cases.

\acknowledgments
This research is supported by grants from the U.S. Department of Energy Office of Science, the U.S.
National Science Foundation and the Smithsonian Institution, by NSERC in Canada, by Science
Foundation Ireland (SFI 10/RFP/AST2748) and by STFC in the U.K. We acknowledge the excellent work
of the technical support staff at the Fred Lawrence Whipple Observatory and at the collaborating
institutions in the construction and operation of the instrument. 

The \textit{Fermi} LAT Collaboration acknowledges generous ongoing support
from a number of agencies and institutes that have supported both the
development and the operation of the LAT as well as scientific data analysis.
These include the National Aeronautics and Space Administration and the
Department of Energy in the United States, the Commissariat \`a l'Energie Atomique
and the Centre National de la Recherche Scientifique / Institut National de Physique
Nucl\'eaire et de Physique des Particules in France, the Agenzia Spaziale Italiana
and the Istituto Nazionale di Fisica Nucleare in Italy, the Ministry of Education,
Culture, Sports, Science and Technology (MEXT), High Energy Accelerator Research
Organization (KEK) and Japan Aerospace Exploration Agency (JAXA) in Japan, and
the K.~A.~Wallenberg Foundation, the Swedish Research Council and the
Swedish National Space Board in Sweden.

Additional support for science analysis during the operations phase is gratefully
acknowledged from the Istituto Nazionale di Astrofisica in Italy and the Centre National d'\'Etudes Spatiales in France.

C.P. gratefully acknowledges 
financial support of the Klaus Tschira Foundation. A.P. acknowledges 
NSF grant AST-0908480 for support.


\bibliographystyle{apj}
\bibliography{refs}

\begin{deluxetable}{lcc}
\tablecolumns{3}
\tablewidth{0pc}
\tablecaption{Regions of interest in the Coma cluster field of view}
\tablehead{
\colhead{Source} &
\colhead{RA (J2000)} &
\colhead{Dec (J2000)}
}
\startdata
Core & \RA{12}{59}{48.7} & \Dec{+27}{58}{50.0}\\
NGC 4889 & \RA{13}{00}{08.13} & \Dec{+27}{58}{37.03}\\
NGC 4874 & \RA{12}{59}{35.71} & \Dec{+27}{57}{33.37}\\
NGC 4921 & \RA{13}{01}{26.12} & \Dec{+27}{53}{09.59}\\
\enddata
\tablecomments{The cluster core is considered both as a point source and a modestly extended
source. Three central galaxies are also considered in point-source searches. The choice is based on
evidence for an excess of nonthermal X-ray emission \citep{article:Neumann_etal:2003} at the
location of these galaxies.}
\label{table:roi}
\end{deluxetable}

\begin{deluxetable}{lccccccccc}
\tablecolumns{10}
\tablewidth{0pc}
\tablecaption{VERITAS VHE gamma-ray flux upper limits for different regions of interest in the
Coma cluster of galaxies and surroundings}
\rotate
\tablehead{
\colhead{Source} & 
\colhead{R\tablenotemark{a} [deg]} & 
\colhead{$N_{S}$\tablenotemark{b}} & 
\colhead{S\tablenotemark{c} [$\sigma$]} &
\multicolumn{6}{c}{Flux U.L.\tablenotemark{d}} \\
\colhead{} & 
\colhead{} & 
\colhead{} & 
\colhead{} &
\multicolumn{2}{c}{$\alpha=2.1$} &
\multicolumn{2}{c}{$\alpha=2.3$} &
\multicolumn{2}{c}{$\alpha=2.5$}
}
\startdata
Core & 0 & 17 & 0.84 & 2.59 & (0.78\%) & 2.78 & (0.83\%) & 2.97 & (0.89\%) \\
& 0.2 & -41 & -1.0 & 1.96 & (0.59\%)  & 2.09 & (0.63\%) & 2.21 & (0.66\%) \\
& 0.4 & -26 & -0.30 & 4.44 & (1.3\%)  & 4.74 & (1.4\%) & 5.02 & (1.5\%)\\
NGC 4889 & 0 & 3 & 0.14 & - & - & 1.85 & (0.55\%)  & - & - \\
NGC 4874 & 0 & -14 & -0.71 & - & - &  1.51 & (0.45\%)  & - & - \\
NGC 4921 & 0 & -4 & -0.23 & - & - &  2.41 & (0.72\%)  & - & -
\enddata
\tablenotetext{a}{Intrinsic source radius (zero means point source), which is convolved with the
gamma-ray point-spread function.}
\tablenotetext{b}{Net event counts in the source region.}
\tablenotetext{c}{Statistical significance calculated according to \citet{article:LiMa:1983}.}
\tablenotetext{d}{99\% confidence level upper limit in units of $10^{-8}$ ph. m$^{-2}$ s$^{-1}$
calculated according to \citet{article:Rolke_etal:2005} above an energy threshold of 220 GeV, with
corresponding fluxes in percent of the steady Crab Nebula flux in parenthesis, for different
values of the spectral index, $\alpha$.}
\label{table:results}
\end{deluxetable}

\begin{deluxetable}{lccc}
\tablecolumns{4}
\tablewidth{0pc}
\tablecaption{\Fermi-LAT HE gamma-ray flux upper limits for the Coma cluster core}
\tablehead{
                      & $1-3$ GeV & $3-10$ GeV & $10-30$ GeV\\
\small{Spatial Model} & \small{Flux U.L.\tablenotemark{a}} (\small{Significance}) & \small{Flux
U.L.\tablenotemark{a}} (\small{Significance}) & \small{Flux U.L.\tablenotemark{a}}
(\small{Significance})\\
\hline
\multicolumn{4}{c}{\small{Spectral index $\alpha=2.1$}}
}
\startdata
\small{Point Source}          & 1.882 (0.000) & 0.759 (0.000) & 0.671 (0.830) \\
\small{Disk: r = 0.2$^\circ$} & 2.109 (0.152) & 0.899 (0.000) & 0.719 (0.740)\\
\small{Disk: r = 0.4$^\circ$} & 2.438 (0.201) & 1.232 (0.619) & 0.875 (1.387)\\
\cutinhead{\small{Spectral index $\alpha=2.3$}}
\small{Point Source}          & 1.946 (0.000) & 0.788 (0.000) & 0.667 (0.874) \\
\small{Disk: r = 0.2$^\circ$} & 2.180 (0.169) & 0.941 (0.000) & 0.725 (0.828)\\
\small{Disk: r = 0.4$^\circ$} & 2.524 (0.246) & 1.275 (0.742) & 0.869 (1.390)\\
\cutinhead{\small{Spectral index $\alpha=2.5$}}
\small{Point Source}          & 2.008 (0.000) & 0.816 (0.000) & 0.663 (0.915)\\
\small{Disk: r = 0.2$^\circ$} & 2.246 (0.189) & 0.979 (0.020) & 0.720 (0.864)\\
\small{Disk: r = 0.4$^\circ$} & 2.606 (0.291) & 1.313 (0.856) & 0.861 (1.387)\\
\enddata
\tablenotetext{a}{99\% confidence level flux upper limit in units of $10^{-6}$ ph. m$^{-2}$ s$^{-1}$.}
\label{table:fermi}
\end{deluxetable}

\begin{deluxetable}{cccccc}
\tablecolumns{6}
\tablewidth{0pc}
\tablecaption{Constraints on the CR-to-thermal pressure ratio in the Coma cluster core (simplified,
isobaric analytic model) for different spatial extensions and predicted fluxes for the energy bands
1-3 GeV and $>220$~GeV (simulation-based model)}
\rotate
\tabletypesize{\small}
\tablehead{
\colhead{R\tablenotemark{a} [deg]} 
& \multicolumn{3}{c}{analytic model: $X_{\CR}$\tablenotemark{b} }
& \colhead{$F_{\gamma,\rmn{sim}}(E)$\tablenotemark{c}}
& \colhead{$F_{\rmn{UL}}/F_{\gamma,\rmn{sim}}(>E)$\tablenotemark{d}} \\
& \colhead{$\alpha=2.1$} & \colhead{$\alpha=2.3$} & \colhead{$\alpha=2.5$} & & 
}
\startdata
\multicolumn{6}{c}{VERITAS constraints}\\
\hline
0   & 0.1   & 0.23 & 0.97 & 1.9 & 14.8\\
0.2 & 0.048 & 0.10 & 0.43 & 2.9 & 7.2\\
0.4 & 0.067 & 0.15 & 0.62 & 4.4 & 10.8\\
\cutinhead{\Fermi constraints}
0   & 0.035 & 0.024 & 0.033 & 1.4 & 1.34\\
0.2 & 0.024 & 0.017 & 0.022 & 2.1 & 1.00\\
0.4 & 0.017 & 0.012 & 0.016 & 3.2 & 0.76\\
\enddata
\tablenotetext{a}{Intrinsic source radius (zero means point source), which is convolved with the
gamma-ray point-spread function.}
\tablenotetext{b}{Constraint on the CR-to-thermal pressure ratio,
$X_{\CR} = \expval{P_\CR}/ \expval{P_\rmn{th}}$, which was assumed to be constant throughout the
cluster and calculated according to \citet{article:PfrommerEnsslin:2004b}.}
\tablenotetext{c}{Integrated gamma-ray flux from the simulation-based analytic model by
\citet{article:PinzkePfrommer:2010}: above $E=220~\rmn{GeV}$ in units of $10^{-9}$ ph. m$^{-2}$
s$^{-1}$ for VERITAS and for $E=1-3~\rmn{GeV}$ in units of $10^{-6}$ ph. m$^{-2}$ s$^{-1}$ for
Fermi.} 
\tablenotetext{d}{Ratio of flux upper limit ($F_{\rmn{UL}}$) to integrated gamma-ray flux
from the simulation-based analytic model, with the U.L. based on spectral index $\alpha=2.3$ in
the VERITAS band ($E>220~\rmn{GeV}$) and $\alpha=2.1$ in the \Fermi band ($E=1-3~\rmn{GeV}$).}
\label{table:constraints_simple}
\end{deluxetable}

\begin{deluxetable}{ccccccc}
\tablecolumns{7}
\tablewidth{0pc}
\tablecaption{Minimum gamma-ray fluxes in the hadronic model of radio halos, where the
radio-emitting electrons are secondaries from CR interactions, and corresponding minimum 
CR-to-thermal pressure ratios for Coma}
\rotate
\tabletypesize{\small}
\tablehead{
\colhead{R\tablenotemark{a} [deg]} 
& \multicolumn{3}{c}{$F_{\gamma,\,\rmn{min}}(>E)$\tablenotemark{b}} 
& \multicolumn{3}{c}{$10^{4}\times X_{\CR,\,\rmn{min}}$\tablenotemark{c} } \\
& \colhead{$\alpha=2.1$} & \colhead{$\alpha=2.3$} & \colhead{$\alpha=2.5$}
& \colhead{$\alpha=2.1$} & \colhead{$\alpha=2.3$} & \colhead{$\alpha=2.5$}
}
\startdata
\multicolumn{7}{c}{VERITAS energy range, $E>220$~GeV}\\
\hline
0 & 1.6 & 0.7 & 0.3 & 6.7 & 6.1 & 11 \\
0.2  & 3.1 & 1.4 & 0.6 & 7.8 & 7.2 & 13 \\
0.4  & 6.3 & 2.8 & 1.3 & 9.8 & 9.0 & 16 \\
\cutinhead{\Fermi energy range, 1--3~GeV}
0 & ~~3.5 & ~~4.8 & ~~6.4 & 6.7 & 6.1 & 11 \\
0.2  & ~~6.8 & ~~9.3 &  12.5 & 7.8 & 7.2 & 13 \\
0.4  &  13.5 & 18.6  &  25.0 & 9.8 & 9.0 & 16 \\
\enddata
\tablenotetext{a}{Intrinsic source radius (zero means point source), which is convolved with the
gamma-ray point-spread function.}
\tablenotetext{b}{Minimum gamma-ray flux derived from the hadronic model described in
\S~\ref{sec:Fmin}. Values are in units of $10^{-10}$ ph. m$^{-2}$ s$^{-1}$ for the VERITAS energy
range and $10^{-8}$ ph. m$^{-2}$ s$^{-1}$ for \Fermi energy range.}
\tablenotetext{c}{Minimum CR-to-thermal pressure ratio, $X_{\CR,\,\rmn{min}}$, in the hadronic model
described in \S~\ref{sec:Fmin}. For simplicity, we duplicate $X_{\CR,\,\rmn{min}}$ for the VERITAS
and \Fermi constraints: for a given realization of the CR pressure (and a magnetic field model
which is trivial here as we assume $B \gg B_\rmn{CMB}$), we can derive the radio flux as well as
gamma-ray fluxes in various bands (1--3 GeV, $>220$ GeV).}
\label{table:constraints}
\end{deluxetable}

\begin{deluxetable}{ccccccc}
\tablecolumns{4} \tablewidth{0pc} 
\tablecaption{Constraints on magnetic fields in the hadronic model of the Coma radio halo and the
corresponding CR-to-thermal pressure ratio (at the largest emission radius of 1 Mpc) such that the model
reproduces the observed radio surface-brightness profile}
\tablehead{ &
\multicolumn{6}{c}{Minimum magnetic field, $B_{0,\rmn{min}}~[\mu\rmn{G}]$\tablenotemark{a}:}\\
& \multicolumn{3}{c}{VERITAS constraints} & \multicolumn{3}{c}{\Fermi constraints} \\
\colhead{$\alpha_{B}$} & \colhead{$\alpha=2.1$} & \colhead{$\alpha=2.3$} &
\colhead{$\alpha=2.5$} & \colhead{$\alpha=2.1$} & \colhead{$\alpha=2.3$} &
\colhead{$\alpha=2.5$} } \startdata
0.3 & 0.69 & 0.57 & 0.48 & 1.38 & 1.95 & 2.68 \\
0.5 & 0.97 & 0.80 & 0.68 & 1.94 & 2.74 & 3.78 \\
0.7 & 1.40 & 1.17 & 0.99 & 2.80 & 3.97 & 5.50 \\
\hline
& \multicolumn{6}{c}{Corresponding $X_{\CR}\, (1\,\rmn{Mpc})$:}\\
\hline
0.3 & 0.46 & 1.05 & ~~4.55 & 0.11 & 0.08 & 0.11 \\
0.5 & 0.74 & 1.70 & ~~7.47 & 0.18 & 0.13 & 0.17 \\
0.7 & 1.09 & 2.59 &  11.55 & 0.27 & 0.19 & 0.26 \\
\enddata
\tablenotetext{a}{The parameters of the magnetic field are the magnetic decline, $\alpha_B$, and
the central field strength, $B_{0}$, which are defined by
$B(r) = B_{0}\,[n_\rmn{e}(r) / n_{\rmn{e}}(0)]^{\alpha_B}$. In all cases, we used the most constraining
$\rmn{R}=0.4^\circ$; see \S~\ref{sec:B} for details.}  
\label{table:Bmin}
\end{deluxetable}

\begin{deluxetable}{ccc}
\tablecolumns{3}
\tablewidth{0pc}
\tablecaption{Astrophysical factors}
\tablehead{
\colhead{R [deg]} &
\colhead{$\expval{J}_{\rmn{signal}}$ [GeV$^{2}$ cm$^{-5}$ sr]} & 
\colhead{$\alpha\expval{J}_{\rmn{bkg}}$ [GeV$^{2}$ cm$^{-5}$ sr]}
}
\startdata
0 & $5.7\times 10^{16}$ & $1.3\times 10^{14}$ (negligible) \\
0.2 & $8.1\times 10^{16}$ & $4.4\times 10^{14}$ ($<0.01\expval{J}_{\rmn{signal}}$, negligible) \\
0.4 & $9.4\times 10^{16}$ & $1.3\times 10^{15}$ ($\simeq0.01\expval{J}_{\rmn{signal}}$, negligible)
\enddata
\tablecomments{$\expval{J}_{\rmn{bkg}}$ is the astrophysical factor calculated for the background
region (ring method) and is used to estimate the level of gamma-ray contamination from DM
annihilation. $\alpha$ is the size ratio of the ON and OFF source regions.}
\label{table:astrofactor}
\end{deluxetable}

\begin{deluxetable}{lccc}
\tablecolumns{4}
\tablewidth{0pc}
\tablecaption{Upper limits on the DM annihilation cross section times velocity $\expval{\sigma v}$
from VERITAS observations of the Coma cluster.}
\tablehead{
\colhead{Channel} & R [deg] & $m_{\chi}$ [GeV] & $\expval{\sigma v}$ [cm$^{3}$ s$^{-1}$]
}
\startdata
W$^{+}$W$^{-}$ & 0 & 2000 & $1.1\times 10^{-20}$ \\
& 0.2 & 1900 & $4.3\times 10^{-21}$ \\
& 0.4 & 1900 & $8.4\times 10^{-21}$ \\
$b\bar{b}$ & 0 & 3500 & $1.2\times 10^{-20}$ \\
& 0.2 & 3400 & $4.4\times 10^{-21}$ \\
& 0.4 & 3500 & $8.7\times 10^{-21}$ \\
$\tau^{+}\tau^{-}$ & 0 & 670 & $2.4\times 10^{-21}$ \\
& 0.2 & 650 & $9.1\times 10^{-22}$ \\
& 0.4 & 660 & $1.8\times 10^{-21}$
\enddata
\tablecomments{Upper limits are shown for different integration regions and DM particle mass
$m_{\chi}$, and are derived from the VERITAS gamma-ray flux upper limits presented in this work.}
\label{table:DMlimits}
\end{deluxetable}

\end{document}